# Derivation of an Algorithm for Calculation of the Intersection Area of a Circle with a Grid with Finite Fill Factor

**Abstract:**

The problem deals with an exact calculation of the intersection area of a circle arbitrary placed on a grid of square shaped elements with gaps between them (finite fill factor). Usually an approximation is used for the calculation of the intersection area of the circle and the squares of the grid. We analyze the geometry of the problem and derive an algorithm for the exact computation of the intersection areas. The results of the analysis are summarized in the tally sheet. In a real world example this might be a CCD or CMOS chip, or the tile structure of a floor.

________

**\*Corresponding Author: Dmitrij Gendler:** Fraunhofer Institute of Optronics, system Technologies and Images Explotation (IOSB), Germany,
E-Mail: dmitrij.gendler@iosb.fraunhofer.de
**Christian Eisele:** Fraunhofer Institute of Optronics, system Technologies and Images Explotation (IOSB), Germany,
E-Mail: christian.eisele@iosb.fraunhofer.de
**Dirk Seiffer:** Fraunhofer Institute of Optronics, system Technologies and Images Explotation (IOSB), Germany,
E-Mail: dirk.seiffer@iosb.fraunhofer.de
**Norbert Wendelstein:** Fraunhofer Institute of Optronics, system Technologies and Images Explotation (IOSB), Germany,
E-Mail: norbert.wendelstein@iosb.fraunhofer.de

**Introduction**

The article includes following parts:

1. **Statement of the Problem**
2. **Base relations and notation**
3. **Constraint relations between geometry and indices of grid**
4. **Preliminary analysis of data**
5. **Some necessary formulae**
6. **Location analysis**
7. **Summary**
8. **References**
9. **Appendix**

All proofs and formulae calculating intersection areas are derived in the "Location analysis" chapter. But, as all results are summarized in the tally sheet, to get the table with all formulae, reader can look directly to "Summary".

To our great surprise we have not found analytical solving of the problem. Therefore all the references are given only into various forums where the problem has been discussed.

1. **Statement of the Problem**

We consider the below given grid:

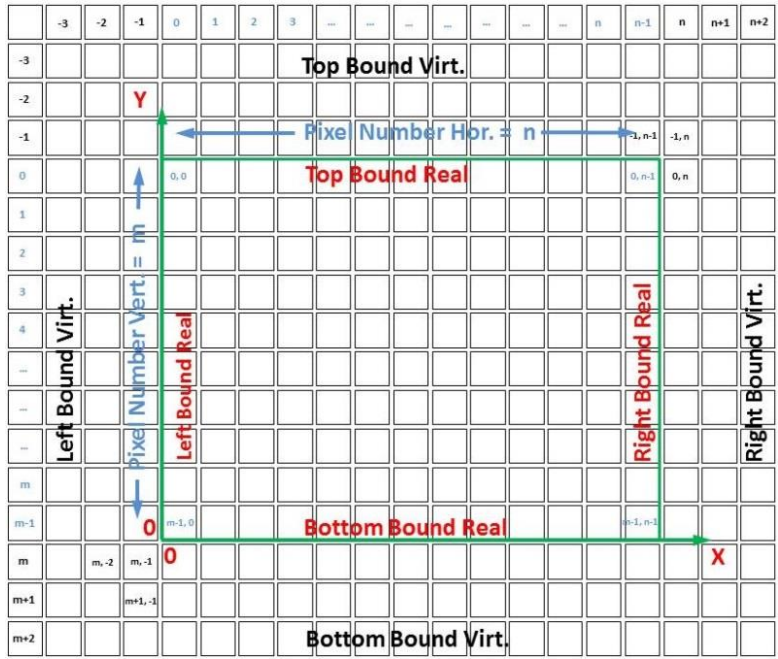

Fig.1. Real and virtual areas of the grid

The problem we are dealing with is the proper calculation of the intersection area of a circle placed on a grid of squares. In a real world example this might be a CCD or CMOS chip, or the tile structure of a floor.

The structure of the grid is as depicted in Fig. 1. Squares with edge length $a$ are arranged in such a way, that the distance between centers of neighboring elements is $b$. $b$ is called the pitch of the grid. The area of one square element is thus $a^2$. The area covered by one square plus its surrounding gap is given by $b^2$. The ratio between the two is called fill factor.

In the following we will distinguish two areas within the grid and call them "real" and "virtual". The real area is the bounded part of endless grid, where intersection square should be computed. The virtual area is a grid out of borders of real grid.

The real area has **m rows** and **n columns** (as indicated by the green line in Fig. 1). Everything lying outside of the green line is the virtual area. Further we shall sometimes use the word "pixel", meaning only squares, which lie in the real area.

Assumed indexing in grid is: from 0 up to m-1 for rows and from 0 up to n-1 for columns of the grid. Correspondingly indices in virtual area are either negative or greater than (m-1) or (n-1).

Taking the assumed indexing into account, the following indices of the corners of the real area are yielded:

Left top         0, 0
Left bottom      0, m-1
Right top        0, n-1
Right bottom     m-1, n-1

Furthermore we will use a Cartesian coordinate system to describe positions of vertices and the circle. The origin of this coordinate system X,Y will be the lower left corner of the green rectangle in Figure 1.

A circle with given radius **R** may be located in an arbitrary place within the grid. We consider its intersections with squares in the underline{real area}, i.e., with pixels.

In order to simplify further calculations we assume that the radius of the circle is at more than twice as large as a side of a square (**R > 2*Size**).

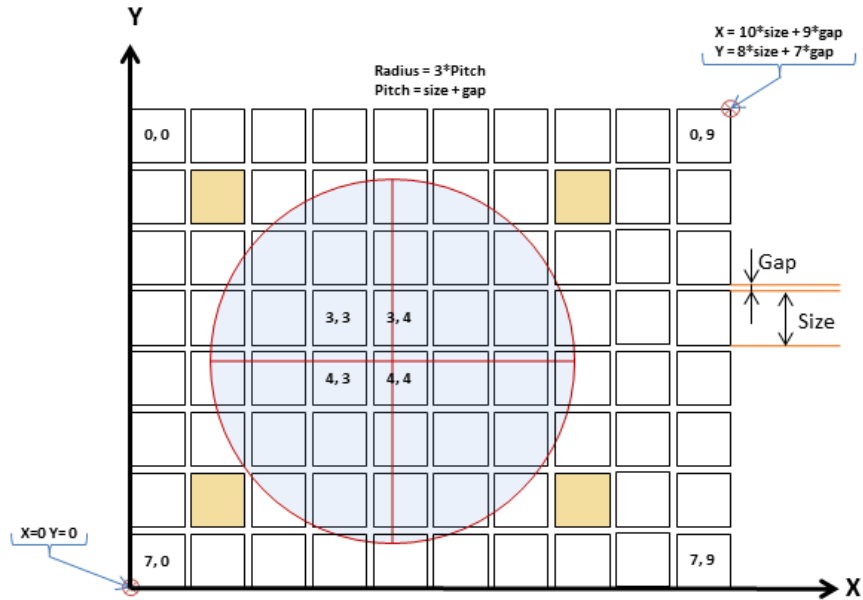

Fig.2. The terms "Size", "Gap" and "Pitch" and their relation with indices

To go further we consider an example setting
**R = 3*Pitch**, where **Pitch = Size + Gap**  (1)

As we are interested only in underline{intersection in the real area} we limit the size of the virtual area to 3 squares + 2 gaps in each direction. Various placing of a circle and intersections are shown below.

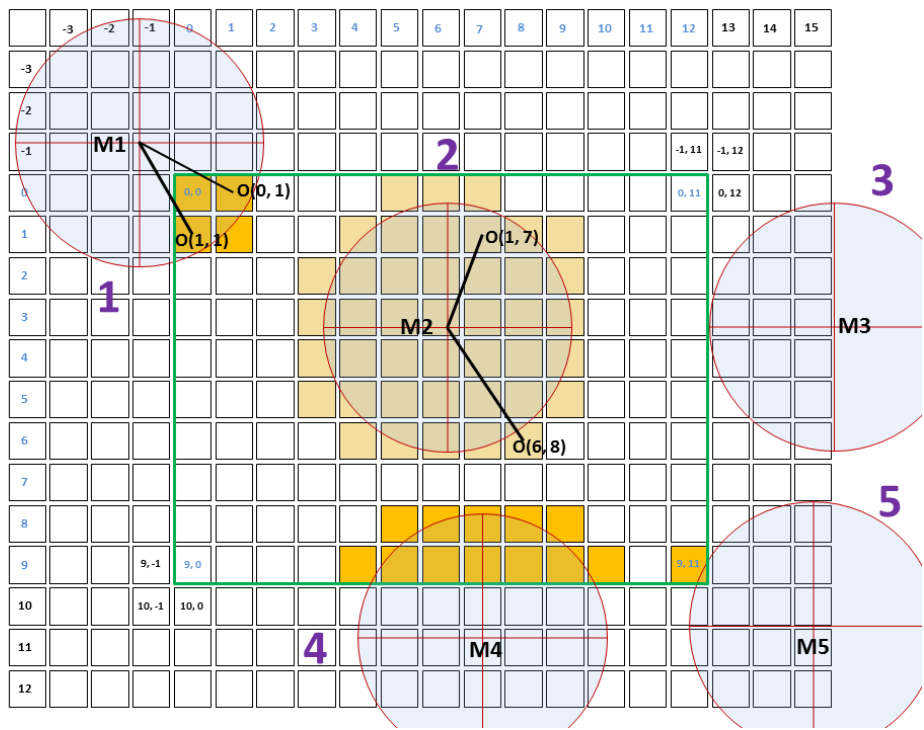

Fig.3. Various types of intersection

The task consists of two parts:

1. Find the list of pixels which are intersected with the circle.
2. Compute the areas of intersection for each member of this list.

## 2. Base Relations and Notation

Recall from above that indices of the pixel matrix lie in these ranges:

Left top        0, 0
Left bottom     0, m-1
Right top       0, n-1
Right bottom    m-1, n-1

From Fig. 1 and (1) we obtain the sizes of the real area as follows.

In horizontal direction (along X axis):   (n-2) * Pitch + Size
In vertical direction (along Y axis):     (m-2) * Pitch + Size

Each square has 4 vertices with coordinate pairs which are denoted $(x_1,y_1)$ ... $(x_4,y_4)$. Analogously the center of the circle has coordinates $(x_0,y_0)$. There is simple relation between indices of a square as an element of the grid and metric coordinates of its vertices in coordinate system XY.

For the square with indices (i, j):

X coordinate of left side       = j*Pitch
X coordinate of right side      = j*Pitch + Size
Y coordinate of top side        = i*Pitch
Y coordinate of bottom side     = i*Pitch + Size

And X, Y coordinate of the pixel corner are:

Left bottom     (j*Pitch, i*Pitch + Size)
Left top        (j*Pitch, i*Pitch)
Right bottom    (j*Pitch + Size, i*Pitch + Size)
Right top       (j*Pitch + Size, i*Pitch)

We may test each pixel whether it is intersected by the circle and create a list containing indices of pixels and the <u>case of intersection</u>.

Distance between the center of the circle and any vertex of a pixel is

$$d_k^2 = \sqrt{(x_k - x_0)^2 + (y_k - y_0)^2} \quad \text{and} \quad d_k^2 = (x_k - x_0)^2 + (y_k - y_0)^2, \quad \text{where } k = 1, 2, 3, 4 \tag{1a}$$

If radius of circle is **R**, then we can get mutual positions of center and vertices as follows:
$R^2 - d^2 < 0$     vertex lies outside the circle     (2a)
$R^2 - d^2 = 0$     vertex lies on the circle     (2b)

$R^2 - d^2 > 0$   vertex lies inside the circle (2c)

Testing of latter relations and additional analysis in some cases is sufficient to create a list of pixels intersected by the circle. But as we are interested not only in obtaining the list of touched pixels, but also in the computation of their areas of intersections, we must provide detailed consideration of possible intersection configurations.

We define the **ordering of vertices** as follows:

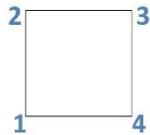

Fig.4. Ordering of vertices

Then, the length of a square side can be computed as follows: Size = $|x_1 - x_4| = |x_2 - x_3| = |y_1 - y_2| = |y_3 - y_4|$
We mention these trivial relations here because further to get length of square side we shall use only the term "Size".

To classify the location of a square relatively to a circle we denote positions of vertices as follows:

0 – vertex lies outside the a circle (3)
1 – vertex lies on the a circle
2 – vertex lies inside the a circle

Then the location of a square in relation to the circle is characterized by an **ordered quadruple**, e.g.,

0000 – square is either fully outside the circle or is intersected
2222 - square is fully inside the circle
0220 – first and fourth vertices are outside the circle, second and third vertices are inside of it
1022 – first vertex is on the circle, second one is outside, third and fourth ones are inside

If we were only interested in the list of concerned pixels, then a member of the list is simply

$$\{i, j\} \quad (4)$$

But if we want to compute the areas of intersection of pixels from the latter list with the circle then additional information is necessary, namely, the quadruple of pixel vertices coordinates and the quadruple of its configuration. A member of this list looks like:

$$\{\{i, j\}, \{(x_1,y_1) \ldots (x_4,y_4)\}, \{v_1 v_2 v_3 v_4\}\}, \quad (5)$$

where   $v_1 \ldots v_4 = \{0, 1, 2\}$   - location of the vertex relative to the circle (6)

## 3. Constraint Relations between Geometry and Indices of Grid

A priori relation between the radius of a circle and the pitch is given widely used in practice looks like:

**R = a * Pitch**,   where **a** – a priori given factor at least 2

Depending on coordinates $(x_0, y_0)$ of a circle's center, it can be computed exactly, which pixels are fully inside a circle and which are intersected.

The proposed method is based on the conservative estimate of the upper bound of pixel array with following selection of array elements according our criterions.

Let the center of a circle falls into the range of the left side of the square (including this side) and the left side of the following square (excluding this side). In other words, center hits either inside the square body or into the gap between two squares. Same considerations are valid for top and bottom sides of squares.

The basic idea of the algorithm is to regard the quadratic area with the square side of about 2*R = 2*a*Pitch with the center, coinciding with the center of the circle $(x_0, y_0)$ and to test all pixels, having at least one vertex or part of the pixel's body inside this quadratic area. In terms of indices this means that all the squares with indices in interval ±N from the indices of hit should be tested. To ensure the approximation of the upper-bound we imply the following rule:

$$N = \lceil 2*a \rceil \quad \text{if } 2*a \text{ is non-integer - ceiling of multiplication}$$
$$N = 2*a + 1 \quad \text{if } 2*a \text{ is integer}$$

E.g., for R = 3*Pitch  a = 3  N = 2 * 3 + 1 = 7
The number of squares to be tested = $N^2$ = 7 * 7 = 49.

If coordinates of a circle's center $(x_0, y_0)$, then (i,j) indices of pixel (or possibly of square in virtual area) of hit can be obtained as follows:

$$i = \left\lfloor \frac{x_0}{\text{Pitch}} \right\rfloor - \text{integer part of quotient} - \text{floor of quotient}$$

Recall that in vertical direction (along the Y axis) size of the real area is (m-2)*Pitch + Size.
In the assumed system the similar relation for j is a little bit trickier:

$$j = \left\lfloor \frac{(m-2)*\text{Pitch} + \text{Size} - y_0}{\text{Pitch}} \right\rfloor - \text{integer part of quotient} - \text{floor of quotient}$$

The fact, that a hit may be in a gap, is unimportant, as we are not trying to obtain an exact solution, but on this step only a good upper-bound constraint.

So, we owe to test all the squares in the area with following combinations of indices:

$$i = \left\lfloor \frac{x_0}{\text{Pitch}} \right\rfloor \pm \lceil a \rceil$$

$$j = \left\lfloor \frac{(m-2)*\text{Pitch} + \text{Size} - y_0}{\text{Pitch}} \right\rfloor \pm \lceil a \rceil$$

Some of so computed indices may not belong to pixels, but to squares, which are inside the virtual area. Therefore, additional tests on validity are necessary. Recalling, that indices in the real area (pixel indices) are bounded to horizontally {0, (n -1)} and vertically {0, (m -1)}, we drop the squares (they are not pixels), indices of which don't satisfy one of two following inequalities:

index i  $\quad 0 <= \left\lfloor \frac{(m-2)*\text{Pitch} + \text{Size} - y_0}{\text{Pitch}} \right\rfloor \pm \lceil a \rceil <= (m - 1)$ \hfill (7)

index j  $\quad 0 <= \left\lfloor \frac{x_0}{\text{Pitch}} \right\rfloor \pm \lceil a \rceil <= (n - 1)$ \hfill (8)

Recall, that we want to get a list with an upper-bounded number of items, each of which represents the coordinates of such pixels, which maybe are intersected with a circle.

E.g., for R = 3*Pitch the minimal number of the intersected pixels is 36. The upper bound in this case is 49. Real number of intersecting pixels depending on the position of the circle's center lies within 36 - 42. This means, that 7 - 13 pixels from the list will be dropped after testing whether they satisfy the inequalities (7) and (8).

## 4. Preliminary Analysis of Data

We are interested in the areas of intersection of pixels and circle. Recalling (3) member of such list looks like

$$\{\{i, j\}, \{(x_1,y_1) \dots (x_4,y_4)\}, \{v_1 v_2 v_3 v_4\}\}, \tag{9}$$

To simplify and clarify further computation we set the system origin to $(x_0, y_0)$, i.e., we set $x_0 = 0$ and $y_0 = 0$ and therefore recalculate for all the members of list (7) the coordinates $(x_1,y_1) \dots (x_4,y_4)$ as follows:

$x_{i\_new} = x_{i\_old} - x_0$

$y_{i\_new} = y_{i\_old} - y_0$

$\{\{i, j\}, \{(x_{1\_old},y_{1\_old}) \dots (x_{4\_old},y_{4\_old})\}, \{v_1 v_2 v_3 v_4\}\} \dashrightarrow \{\{i, j\}, \{(x_{1\_new},y_{1\_new}) \dots (x_{4\_new},y_{4\_new})\}, \{v_1 v_2 v_3 v_4\}\}$ (10)

Obviously formulae for distance between the center of a circle and any vertex of a pixel changes as well:

$$d_k^2 = \sqrt{x_k^2 + y_k^2} \quad \text{and} \quad d_k^2 = x_k^2 + y_k^2, \quad \text{where k = 1, 2, 3, 4} \tag{10a}$$

Recalling assumed notation all possible locations of squares relatively a circle (not taking into account right ordering) are: 0000, 0001, 0002, 0011, 0012, 0022, 0112, 0122, 0222, 1122, 1222, 2222.

Three latter cases are trivial when squares are fully inside a circle

$$S_{1222} = S_{1122} = S_{2222} = Size^2 \tag{11}$$

We introduce the following notation:
$$Z_1 = x_1^2 + y_1^2$$
$$Z_2 = x_2^2 + y_2^2$$
$$Z_3 = x_3^2 + y_3^2$$
$$Z_4 = x_4^2 + y_4^2$$
(12)

$P = R^2$

Obviously
| | | |
|---|---|---|
| $Z_i > P$ | - vertex lies outside a circle | 0 |
| $Z_i = P$ | - vertex lies on a circle | 1 |
| $Z_i < P$ | - vertex lies inside a circle | 2 |

(13)

Accordingly all possible relations are as follows:

$Z_1 + Z_3 > 2*P \qquad Z_2 + Z_4 > 2*P$ (14)
$Z_1 + Z_3 = 2*P \qquad Z_2 + Z_4 = 2*P$
$Z_1 + Z_3 < 2*P \qquad Z_2 + Z_4 < 2*P$

As $\qquad x_2 = x_1 \qquad\qquad x_4 = x_3$ (15)

$$y_2 = y_3 \qquad\qquad y_4 = y_1$$

Note, that

$$Z_2 + Z_4 = (x_2^2 + y_2^2) + (x_4^2 + y_4^2) = (x_1^2 + y_3^2) + (x_3^2 + y_1^2) = (x_1^2 + y_1^2) + (x_3^2 + y_3^2) = Z_1 + Z_3$$

$$Z_2 + Z_4 = Z_1 + Z_3 \tag{16}$$

As it is sufficient to use **only indices 1 and 3** to provide an analysis, from now on we use only these two indices while considering x and y.

Taking relations (10) into account and suppressing the word "new", we get the list (5) in the following form and recalling the assumed ordering (figure 6)

$$\{\{i,j\}, \{(x_1, y_1), (x_2, y_2), (x_3, y_3), (x_4, y_4)\}, \{v_1 v_2 v_3 v_4\}\} \qquad \text{using indices 1 ... 4}$$
$$\{\{i,j\}, \{(x_1, y_1), (\mathbf{x_1, y_3}), (x_3, y_3), (\mathbf{x_3, y_1})\}, \{v_1 v_2 v_3 v_4\}\} \qquad \text{from now on} \tag{17}$$

Not all possible combinations of 0, 1 and 2 forming location vector $\{v_1 v_2 v_3 v_4\}$ correspond to some real square locations. This makes additional constraints which should be taken into account and will be considered later.

Assuming all abovementioned reasons the sought-for algorithm looks like

1) compute expressions for minimal and maximal indices **i** and **j** from formulae (7) and (8);
2) create a list $\{i, j\}$, where **i** and **j** range over the values $i_{min}$ - $i_{max}$, $j_{min}$ - $j_{max}$;
3) for each member of latter list test the relations (2a – 2c) and create the vector $\mathbf{v_1 v_2 v_3 v_4}$;
4) if $\qquad v_1 + v_2 + v_3 + v_4 \leq 1 \qquad$ test whether following conditions holds
$$(|x_1| \geq R) \,\&\&\, (|x_3| \geq R) \,\&\&\, (|y_1| \geq R) \,\&\&\, (|y_3| \geq R) \tag{18}$$
5) if relation (18) holds exclude the member from the list.

$$\{\{i,j\}, \{(x_1, y_1), (x_1, y_3), (x_3, y_3), (x_3, y_1)\}, \{v_1 v_2 v_3 v_4\}\} \qquad \text{reduced list (17)} \tag{19}$$

The reduced list (19) contains only such pixels which are really overlapped by a circle. Moreover, the list contains only two indices 1 and 3 instead of the four indices 1, 2, 3, 4 to simplify the analysis.
Before proceeding to the analysis of various intersections of the circle with the pixels, recall some geometric formulae, needed for further calculations.

## 5. Some Necessary Formulae

Depending on the mutual location of a circle and of a pixel the intersection area can have different geometrical form. Further we shall consider all the possible mutual locations and arising geometrical forms.

It will be shown later, that any of these forms consists of the sum of the standard figures such as the circular segment, the trapezoid, the triangle and the rectangle.

It is obvious that formulae for the areas of these standard figures depending on the coordinates of the pixel vertices should be very similar. <u>Nevertheless, we have deliberately refused from parameterization of the similar formulae to preserve the clarity.</u>

However, some formulae, namely for the circular segment and its chord we present here because they will be used everywhere.

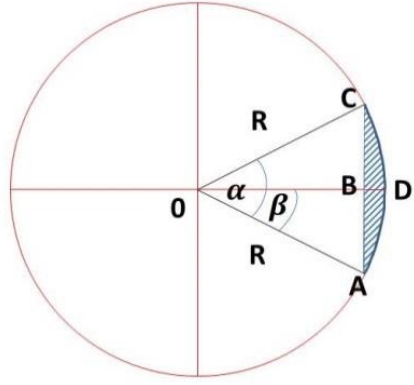

Fig.5. Calculation of the area of the segment ABCD and of length of the chord AC

The chord AC length is $\quad 2*R*\sin\frac{\alpha}{2} = 2*R*\sin\beta$

The semi-chord AB length is $\quad R*\sin\frac{\alpha}{2} = R*\sin\beta$

The area of the circular segment ABCD is $\quad \frac{R^2}{2} * (\alpha - \sin\alpha)$

$\sin\beta = \frac{AB}{R}$

The angle $\beta$ is $\quad \arcsin\left(\frac{AB}{R}\right)$

The angle $\alpha$ is $\quad 2*\beta = 2*\arcsin\left(\frac{AB}{R}\right)$

$\sin\alpha = \sin(2*\beta) = \sin\left[2*\arcsin\left(\frac{AB}{R}\right)\right] = 2*\sin\left[\arcsin\left(\frac{AB}{R}\right)\right]*\cos\left[\arcsin\left(\frac{AB}{R}\right)\right] = 2*\left(\frac{AB}{R}\right)*\sqrt{1-\left(\frac{AB}{R}\right)^2}$

Thus

$$S_{segmABCD} = \frac{R^2}{2}*\left[2*\arcsin\left(\frac{AB}{R}\right) - 2*\left(\frac{AB}{R}\right)*\sqrt{1-\left(\frac{AB}{R}\right)^2}\right] = R^2*\left[\arcsin\left(\frac{AB}{R}\right) - \left(\frac{AB}{R}\right)*\sqrt{1-\left(\frac{AB}{R}\right)^2}\right]$$

Using the length of chord AC:

$$S_{segmABCD} = R^2*\left[\arcsin\left(\frac{AC}{2R}\right) - \left(\frac{AC}{2R}\right)*\sqrt{1-\left(\frac{AC}{2R}\right)^2}\right]$$

From now on for simplicity and brevity we denote the two latter formulae $S_{segm}(AB)$ and $S_{segm}(AC)$ (20)
The length of chord AC will be used as follows:

$AC = \sqrt{(x_a - x_c)^2 + (y_a - y_c)^2} = F(x_a, x_c, y_a, y_c)$ (21)

$S_{segmABCD} = S_{segm}(AC) = S_{segm}(F(x_a, x_c, y_a, y_c))$ (22)

We shall frequently compute the value $\sqrt{R^2 - u^2}$. Therefore denote

$$\bar{u} = \sqrt{R^2 - u^2}$$ (23)

and use it everywhere.

## 6. Location analysis

### 6.1. Four vertices outside

$$S_{segmABCD} = S_{segm}(AC)$$

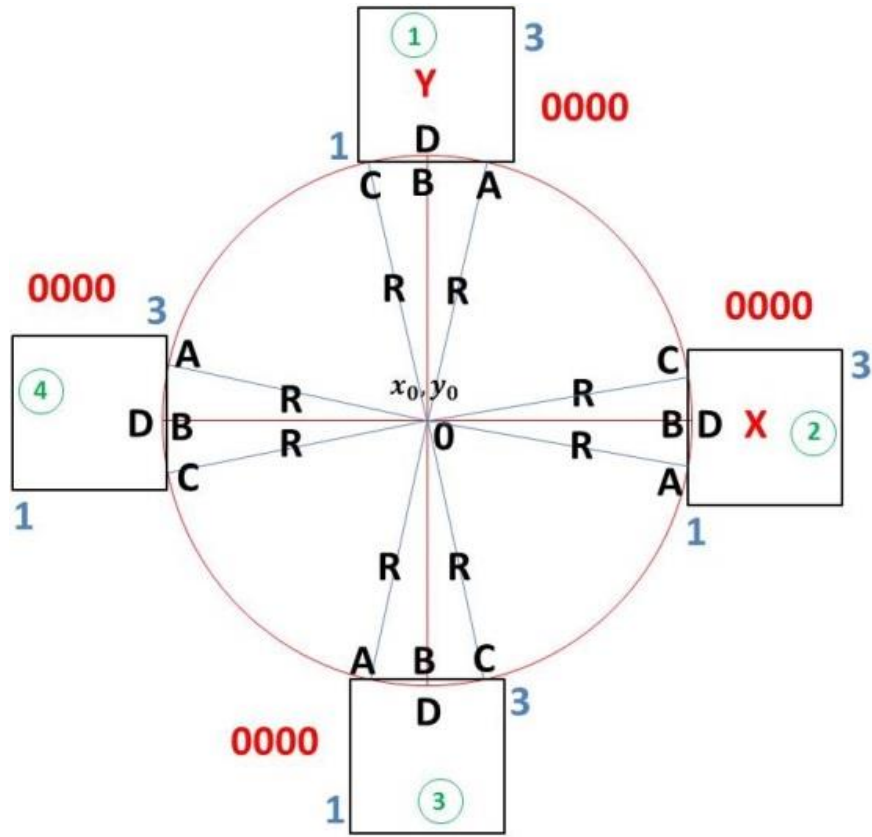

Fig.6. Location "Four vertices outside"

Although all vertices are outside a circle intersection is yet possible, if distance between center of a circle and one of sides of square is less than radius. So we must check holding one of conditions in below given table.

As an example we consider case 1.
Obviously the triangle AOC is isosceles. Therefore  AC = 2 * AB = 2 * $\overline{y_a}$ = 2 * $\overline{y_1}$
Analogously reasoning for all cases is given in the table below.

**Table.1. Conditions and Arguments for the Location "Four vertices outside"**

|   | Location | Condition | AC |
|---|---|---|---|
| 1 | 0000 | $|y_1| < R < |y_3|$ | 2 * $\overline{y_1}$ |
| 2 | 0000 | $|x_1| < R < |x_3|$ | 2 * $\overline{x_1}$ |
| 3 | 0000 | $|y_3| < R < |y_1|$ | 2 * $\overline{y_3}$ |
| 4 | 0000 | $|x_3| < R < |x_1|$ | 2 * $\overline{x_3}$ |

$$S_{segmABCD} = S_{segm}(AC)$$

## 6.2. One vertex on a circle, three outside

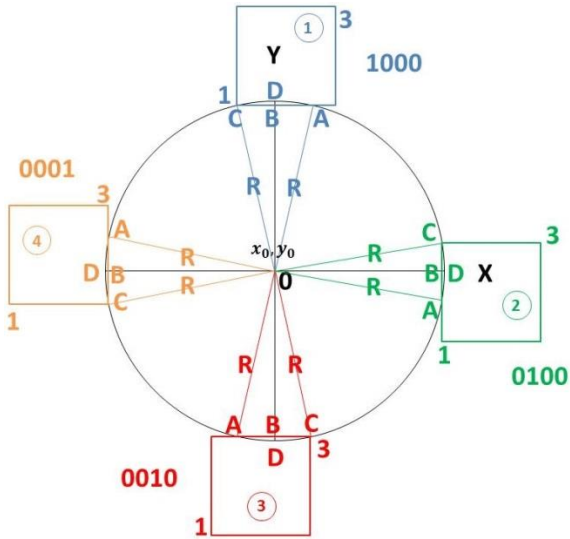

$$S_{segmABCD} = S_{segm}(AC)$$

We outline various locations of squares with color.

As an example we consider case 1.
Obviously the triangle AOC is isosceles. Therefore
$$AC = 2 * CB = 2 * \overline{y_c} = 2 * \overline{y_1}$$

Analogously reasoning for all cases is given in the table below.

Fig.7. Location "One vertex on a circle, three outside"

Table.2. Conditions and Arguments for the Location "One vertex on a circle, three outside"

|   | Location | Condition | AC |
|---|---|---|---|
| 1 | 1000 | $\|y_1\| < R < \|y_3\|$ | $2 * \overline{y_1}$ |
| 2 | 0100 | $\|x_1\| < R < \|x_3\|$ | $2 * \overline{x_1}$ |
| 3 | 0010 | $\|y_3\| < R < \|y_1\|$ | $2 * \overline{y_3}$ |
| 4 | 0001 | $\|x_3\| < R < \|x_1\|$ | $2 * \overline{x_3}$ |

## 6.3. One vertex inside, three vertices outside

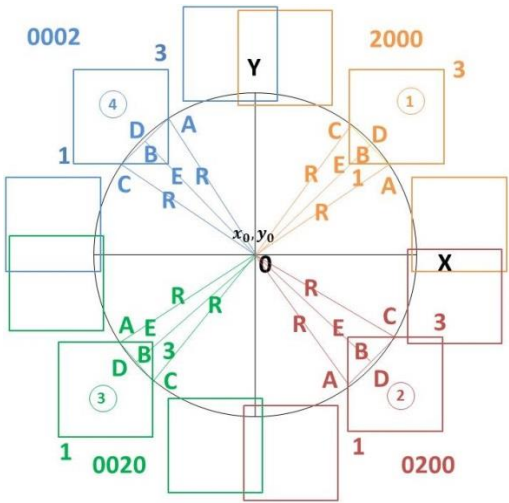

$$S_{intersection} = S_{segmABCD} + S_{triangleAEC}$$

$$S_{segmABCD} = S_{segm}(AC)$$

$$S_{triangleAEC} = \frac{AE * EC}{2}$$

We outline various locations of squares with color and use formula

$$AC = \sqrt{AE^2 + EC^2}$$

Fig.8. Location "One vertex inside, three vertices outside"

Table.3. Conditions and Arguments for the Location "One vertex inside, three vertices outside"

|   | Location | AE | | EC | | AC |
|---|---|---|---|---|---|---|
| 1 | 2000 | $\|x_a - x_c\|$ | $\|\overline{y_1} - x_1\|$ | $\|y_c - y_a\|$ | $\|\overline{x_1} - y_1\|$ | $\sqrt{\|\overline{y_1} - x_1\|^2 + \|\overline{x_1} - y_1\|^2}$ |
| 2 | 0200 | $\|y_c - y_a\|$ | $\|\overline{x_1} - y_3\|$ | $\|x_a - x_c\|$ | $\|\overline{y_3} - x_1\|$ | $\sqrt{\|\overline{x_1} - y_3\|^2 + \|\overline{y_3} - x_1\|^2}$ |
| 3 | 0020 | $\|x_a - x_c\|$ | $\|\overline{y_3} - x_3\|$ | $\|y_c - y_a\|$ | $\|\overline{x_3} - y_3\|$ | $\sqrt{\|\overline{y_3} - x_3\|^2 + \|\overline{x_3} - y_3\|^2}$ |
| 4 | 0002 | $\|y_c - y_a\|$ | $\|\overline{x_3} - y_1\|$ | $\|x_a - x_c\|$ | $\|\overline{y_1} - x_3\|$ | $\sqrt{\|\overline{x_3} - y_1\|^2 + \|\overline{y_1} - x_3\|^2}$ |

Result table in compact form:

Table.4. Conditions and Arguments and for the Location "One vertex inside, three vertices outside". Compact Form

| | Location | Arguments of F | $S_{triangleAEC}$ | AC |
|---|---|---|---|---|
| 1 | 2000 | $\overline{y_1}, x_1, y_1, \overline{x_1}$ | $\dfrac{|\overline{y_1} - x_1| * |\overline{x_1} - y_1|}{2}$ | $\sqrt{|\overline{y_1} - x_1|^2 + |\overline{x_1} - y_1|^2}$ |
| 2 | 0200 | $\overline{y_3}, x_1, y_3, \overline{x_1}$ | $\dfrac{|\overline{x_1} - y_3| * |\overline{y_3} - x_1|}{2}$ | $\sqrt{|\overline{x_1} - y_3|^2 + |\overline{y_3} - x_1|^2}$ |
| 3 | 0020 | $\overline{y_3}, x_3, y_3, \overline{x_3}$ | $\dfrac{|\overline{y_3} - x_3| * |\overline{x_3} - y_3|}{2}$ | $\sqrt{|\overline{y_3} - x_3|^2 + |\overline{x_3} - y_3|^2}$ |
| 4 | 0002 | $\overline{y_1}, x_3, \overline{x_3}, y_1$ | $\dfrac{|\overline{x_3} - y_1| * |\overline{y_1} - x_3|}{2}$ | $\sqrt{|\overline{x_3} - y_1|^2 + |\overline{y_1} - x_3|^2}$ |

### 6.4. Two vertices on a circle, two outside

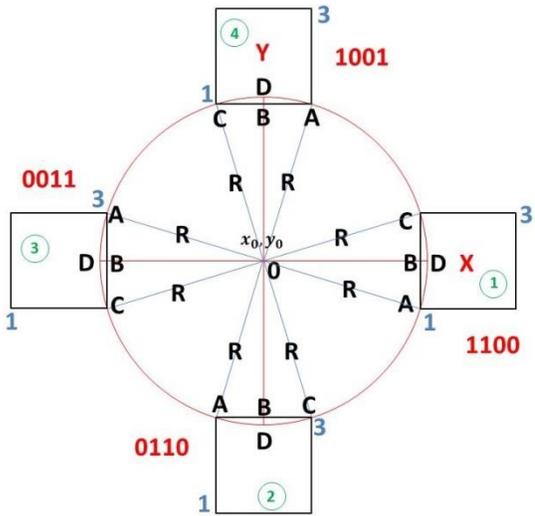

$S_{segmABCD} = S_{segm}(AC)$

Obviously AC = Size

Table.5. Conditions and Arguments for the Location "Two vertices on a circle, two outside"

| | Location | AC |
|---|---|---|
| 1 | 1100 | Size |
| 2 | 0110 | |
| 3 | 0011 | |
| 4 | 1001 | |

Fig.9. Location "Two vertices on a circle, two outside"

Locations of squares represented at the above figure don't exhaust all possible four-digit combinations of two "0" and two "1". We check all possible cases in order not to lose some of them. So, all the cases are:

0011   0110   1001   1100   1010   0101

As first 4 ones are presented above we must check only two cases:   1010 and 0101.

Recall the formulae (12 - 16) and consider the combination 1010 which means, that

$Z_1 = P$ $\qquad\qquad Z_2 > P$
$Z_3 = P$ $\qquad\qquad Z_4 > P$

$Z_1 + Z_3 = 2*P$ $\qquad\qquad Z_2 + Z_4 > 2*P$

As $\quad Z_2 + Z_4 = Z_1 + Z_3 > 2*P$
and $\qquad\quad Z_1 + Z_3 = 2*P$

Latter contradiction means, that combination 1010 corresponds to no location.

Analogously consider the combination 0101 which means, that

$Z_1 > P$                      $Z_2 = P$
$Z_3 > P$                      $Z_4 = P$

$Z_1 + Z_3 > 2*P$           $Z_2 + Z_4 = 2*P$

As     $Z_2 + Z_4 = Z_1 + Z_3 = 2*P$
and           $Z_1 + Z_3 > 2*P$

Latter contradiction means, that combination 0101 corresponds to no location.

### 6.5. One vertex inside, one vertex on a circle, two vertices outside

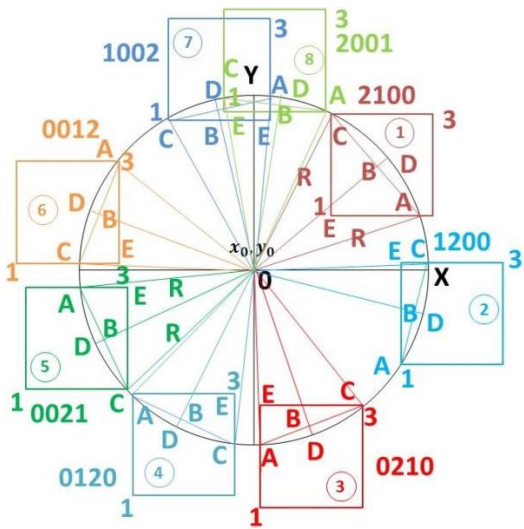

$$S_{intersection} = S_{segmABCD} + S_{triangleAEC}$$

$$S_{segmABCD} = S_{segm}(AC)$$

$$S_{triangleAEC} = \frac{AE * EC}{2}$$

Fig.10. Location "One vertex inside, one vertex on a circle, two vertices outside"

We outline various locations of squares with color and use formula     $AC = \sqrt{AE^2 + EC^2}$

Table.6. Conditions and Arguments for the Location "One vertex inside, one vertex on a circle, two vertices outside"

|   | Location | $x_a$ | $x_c$ | $y_a$ | $y_c$ | AE | | EC | | AC |
|---|---|---|---|---|---|---|---|---|---|---|
| 1 | 2100 | $\overline{y_1}$ | $x_1$ | $y_1$ | $y_3$ | $|x_a - x_c|$ | $|\overline{y_1} - x_1|$ | Size | Size | $\sqrt{|\overline{y_1} - x_1|^2 + Size^2}$ |
| 2 | 1200 | $x_1$ | $\overline{y_3}$ | $y_1$ | $y_3$ | Size | Size | $|x_c - x_a|$ | $|\overline{y_3} - x_1|$ | $\sqrt{Size^2 + |\overline{y_3} - x_1|^2}$ |
| 3 | 0210 | $x_1$ | $x_3$ | $\overline{x_1}$ | $y_3$ | $|y_c - y_a|$ | $|\overline{x_3} - y_3|$ | Size | Size | $\sqrt{|\overline{x_3} - y_3|^2 + Size^2}$ |
| 4 | 0120 | $x_1$ | $x_3$ | $y_3$ | $\overline{x_3}$ | Size | Size | $|y_a - y_c|$ | $|\overline{x_3} - y_1|$ | $\sqrt{Size^2 + |\overline{x_3} - y_1|^2}$ |
| 5 | 0021 | $\overline{y_3}$ | $x_3$ | $y_3$ | $y_1$ | $|x_a - x_c|$ | $|\overline{y_3} - x_3|$ | Size | Size | $\sqrt{|\overline{y_3} - x_3|^2 + Size^2}$ |
| 6 | 0012 | $x_3$ | $\overline{y_1}$ | $y_3$ | $y_1$ | Size | Size | $|x_c - x_a|$ | $|\overline{y_1} - x_3|$ | $\sqrt{Size^2 + |\overline{y_1} - x_3|^2}$ |
| 7 | 1002 | $x_3$ | $x_1$ | $\overline{x_3}$ | $y_1$ | $|y_c - y_a|$ | $|\overline{x_3} - y_1|$ | Size | Size | $\sqrt{|\overline{x_3} - y_1|^2 + Size^2}$ |
| 8 | 2001 | $x_3$ | $x_1$ | $y_1$ | $\overline{x_1}$ | Size | Size | $|y_a - y_c|$ | $|\overline{x_1} - y_1|$ | $\sqrt{Size^2 + |\overline{x_1} - y_1|^2}$ |

Result table in compact form:

Table.7. Conditions and Arguments for the Location "One vertex inside, one vertex on a circle, two vertices outside". Compact Form

| | Location | Arguments of F | $S_{triangleAEC}$ | AC |
|---|---|---|---|---|
| 1 | 2100 | $\overline{y_1}, x_1, y_1, y_3$ | $\dfrac{|\overline{y_1} - x_1| * \text{Size}}{2}$ | $\sqrt{|\overline{y_1} - x_1|^2 + \text{Size}^2}$ |
| 2 | 1200 | $x_1, \overline{y_3}, y_1, y_3$ | $\dfrac{|\overline{y_3} - x_1| * \text{Size}}{2}$ | $\sqrt{|\overline{y_3} - x_1|^2 + \text{Size}^2}$ |
| 3 | 0210 | $x_1, x_3, \overline{x_1}, y_3$ | $\dfrac{|\overline{x_3} - y_3| * \text{Size}}{2}$ | $\sqrt{|\overline{x_3} - y_3|^2 + \text{Size}^2}$ |
| 4 | 0120 | $x_1, x_3, y_3, \overline{x_3}$ | $\dfrac{|\overline{x_3} - y_1| * \text{Size}}{2}$ | $\sqrt{|\overline{x_3} - y_1|^2 + \text{Size}^2}$ |
| 5 | 0021 | $\overline{y_3}, x_3, y_3, y_1$ | $\dfrac{|\overline{y_3} - x_3| * \text{Size}}{2}$ | $\sqrt{|\overline{y_3} - x_3|^2 + \text{Size}^2}$ |
| 6 | 0012 | $x_3, \overline{y_1}, y_3, y_1$ | $\dfrac{|\overline{y_1} - x_3| * \text{Size}}{2}$ | $\sqrt{|\overline{y_1} - x_3|^2 + \text{Size}^2}$ |
| 7 | 1002 | $x_3, x_1, \overline{x_3}, y_1$ | $\dfrac{|\overline{x_3} - y_1| * \text{Size}}{2}$ | $\sqrt{|\overline{x_3} - y_1|^2 + \text{Size}^2}$ |
| 8 | 2001 | $x_3, x_1, y_1, \overline{x_1}$ | $\dfrac{|\overline{x_1} - y_1| * \text{Size}}{2}$ | $\sqrt{|\overline{x_1} - y_1|^2 + \text{Size}^2}$ |

Combinations 0102, 0201, 1020, 2010 are absent in the table 7 because they correspond to no locations. Below we prove this assertion. Recall the formulae (12 - 16).

**Case 0102**

$Z_1 > P$        $Z_2 = P$
$Z_3 > P$        $Z_4 < P$

$Z_1 + Z_3 > 2*P$        $Z_2 + Z_4 < 2*P$

As      $Z_2 + Z_4 = Z_1 + Z_3 < 2*P$
and            $Z_1 + Z_3 > 2*P$

Latter contradiction means, that combination 0102 corresponds to no location.

**Case 0201**

$Z_1 > P$        $Z_2 < P$
$Z_3 > P$        $Z_4 = P$

$Z_1 + Z_3 > 2*P$        $Z_2 + Z_4 < 2*P$

As      $Z_2 + Z_4 = Z_1 + Z_3 < 2*P$
and            $Z_1 + Z_3 > 2*P$

Latter contradiction means, that combination 0201 corresponds to no location.

**Case 1020**

$Z_1 = P$        $Z_2 > P$
$Z_3 < P$        $Z_4 > P$

$Z_1 + Z_3 < 2*P$        $Z_2 + Z_4 > 2*P$

As       $Z_2 + Z_4 = Z_1 + Z_3 > 2*P$
and       $Z_1 + Z_3 < 2*P$

Latter contradiction means, that combination 1020 corresponds to no location.

**Case 2010**

$Z_1 < P$                           $Z_2 > P$
$Z_3 = P$                           $Z_4 > P$

$Z_1 + Z_3 < 2*P$                   $Z_2 + Z_4 > 2*P$

As       $Z_2 + Z_4 = Z_1 + Z_3 > 2*P$
and       $Z_1 + Z_3 < 2*P$

Latter contradiction means, that combination 2010 corresponds to no location.

### 6.6. Two vertices inside, two vertices outside

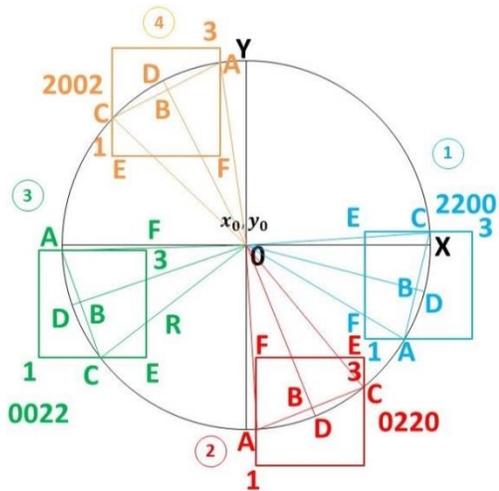

$S_{intersection} = S_{segmABCD} + S_{trapezoidAFEC}$

$S_{segmABCD} = S_{segm}(AC)$

$S_{trapezoidAFEC} = \frac{AF+ EC}{2} * Size$

Fig.11. Location "Two vertices inside, two vertices outside"

We outline various locations of squares with color and use formula

$AC = \sqrt{(x_a - x_c)^2 + (y_a - y_c)^2}$

Table.8. Conditions and Arguments for the Location "Two vertices inside, two vertices outside"

| | Location | $x_a$ | $x_c$ | $y_a$ | $y_c$ | AF | | EC | | $S_{trapezoidAFEC}$ |
|---|---|---|---|---|---|---|---|---|---|---|
| 1 | 2200 | $\overline{y_1}$ | $\overline{y_3}$ | $y_1$ | $y_3$ | $\|x_a - x_1\|$ | $\|\overline{y_1} - x_1\|$ | $\|x_c - x_1\|$ | $\|\overline{y_3} - x_1\|$ | $(\|\overline{y_1} - x_1\| + \|\overline{y_3} - x_1\|) * \frac{Size}{2}$ |
| 2 | 0220 | $x_1$ | $x_3$ | $\overline{x_1}$ | $\overline{x_3}$ | $\|y_a - y_3\|$ | $\|\overline{x_1} - y_3\|$ | $\|y_c - y_3\|$ | $\|\overline{x_3} - y_3\|$ | $(\|\overline{x_1} - y_3\| + \|\overline{x_3} - y_3\|) * \frac{Size}{2}$ |
| 3 | 0022 | $\overline{y_3}$ | $\overline{y_1}$ | $y_3$ | $y_1$ | $\|x_a - x_3\|$ | $\|\overline{y_3} - x_3\|$ | $\|x_c - x_3\|$ | $\|\overline{y_1} - x_3\|$ | $(\|\overline{y_3} - x_3\| + \|\overline{y_1} - x_3\|) * \frac{Size}{2}$ |
| 4 | 2002 | $x_3$ | $x_1$ | $\overline{x_3}$ | $\overline{x_1}$ | $\|y_a - y_1\|$ | $\|\overline{x_3} - y_1\|$ | $\|y_c - y_1\|$ | $\|\overline{x_1} - y_1\|$ | $(\|\overline{x_3} - y_1\| + \|\overline{x_1} - y_1\|) * \frac{Size}{2}$ |

Result table in compact form:

Table.9. Conditions and Arguments for the Location "Two vertices inside, two vertices outside". Compact Form

| | Location | Arguments of F | AC | $S_{trapezoidAFEC}$ |
|---|---|---|---|---|
| 1 | 2200 | $\overline{y_1}, \overline{y_3}, y_1, y_3$ | $\sqrt{(\overline{y_1} - \overline{y_3})^2 + (y_1 - y_3)^2}$ | $(|\overline{y_1} - x_1| + |\overline{y_3} - x_1|) * \frac{Size}{2}$ |
| 2 | 0220 | $x_1, x_3, \overline{x_1}, \overline{x_3}$ | $\sqrt{(x_1 - x_3)^2 + (\overline{x_1} - \overline{x_3})^2}$ | $(|\overline{x_1} - y_3| + |\overline{x_3} - y_3|) * \frac{Size}{2}$ |
| 3 | 0022 | $\overline{y_3}, \overline{y_1}, y_3, y_1$ | $\sqrt{(\overline{y_3} - \overline{y_1})^2 + (y_3 - y_1)^2}$ | $(|\overline{y_3} - x_3| + |\overline{y_1} - x_3|) * \frac{Size}{2}$ |
| 4 | 2002 | $x_3, x_1, \overline{x_3}, \overline{x_1}$ | $\sqrt{(x_3 - x_1)^2 + (\overline{x_3} - \overline{x_1})^2}$ | $(|\overline{x_3} - y_1| + |\overline{x_1} - y_1|) * \frac{Size}{2}$ |

Show that combinations 0202, 2020 correspond to no locations. Recall formulae (12 - 16).

**Case 0202**

$Z_1 > P$  $\qquad$ $Z_2 < P$
$Z_3 > P$  $\qquad$ $Z_4 < P$

$Z_1 + Z_3 > 2*P$ $\qquad$ $Z_2 + Z_4 < 2*P$

As $\quad Z_2 + Z_4 = Z_1 + Z_3 < 2*P$
and $\quad\quad\quad Z_1 + Z_3 > 2*P$
Latter contradiction means, that combination 0102 corresponds to no location.

**Case 2020**

$Z_1 < P$ $\qquad$ $Z_2 > P$
$Z_3 < P$ $\qquad$ $Z_4 > P$

$Z_1 + Z_3 < 2*P$ $\qquad$ $Z_2 + Z_4 > 2*P$

As $\quad Z_2 + Z_4 = Z_1 + Z_3 > 2*P$
and $\quad\quad\quad Z_1 + Z_3 < 2*P$
Latter contradiction means, that combination 0102 corresponds to no location.

**There is one special case when circumference intersects a square in four points.**

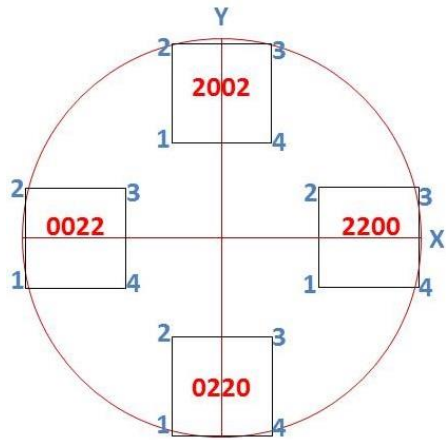

Fig.11a. Location "Two vertices inside, two vertices outside"

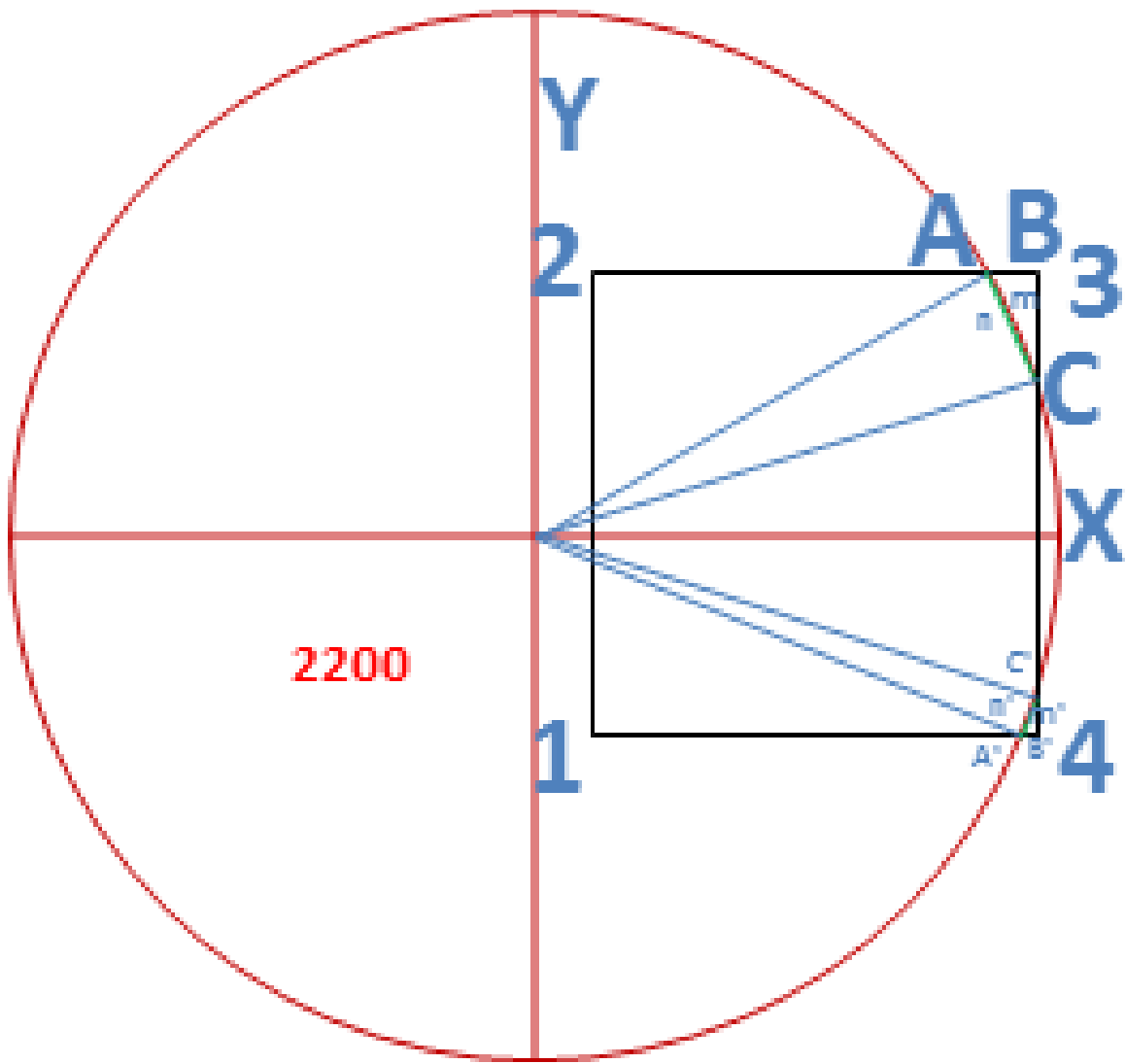

Fig.11b. Special case "Enlarged image 2200"

$$S_{intersection} = S_{square} - S_{triangleABC} + S_{segmAmCn} - S_{triangleA'B'C'} + S_{segmA'm'C'n'}$$

$$S_{segmAmCn} = S_{segm}(AC) \qquad S_{segmA'm'C'n'} = S_{segm}(A'C')$$

$$S_{triangleABC} = \frac{AB * BC}{2} = \frac{(x_b - x_a) * (y_b - y_c)}{2} = \frac{(x_3 - \overline{y_3}) * (y_3 - \overline{x_3})}{2}$$

$$S_{triangleA'B'C'} = \frac{A'B' * B'C'}{2} = \frac{(x_{b'} - x_{a'}) * (y_{c'} - y_{b'})}{2} = \frac{(x_3 - \overline{x_3}) * (\overline{x_3} - y_1)}{2}$$

$$AC = F(x_a, x_c, y_a, y_c) = F(\overline{y_3}, x_3, y_3, \overline{x_3}) \qquad A'C' = F'(x_{a'}, x_{c'}, y_{a'}, y_{c'}) = F'(\overline{y_1}, x_3, y_1, \overline{x_3})$$

Rest 3 cases should be calculated analogously. We don't show here other big pictures and give the results in the below table. All three other cases are presented in the Appendix.

**Table.9a. Conditions and Arguments for the Location "Special case. Two vertices inside, two vertices outside. 4 intersections".**

| | Special Case | Condition | Args of F - AC | Args of F - A´C´ | $S_{segm}(AC)$ | $S_{segm}(A´C´)$ | $S_{triABC}$ | $S_{triA´B´C´}$ | $S_{square}$ |
|---|---|---|---|---|---|---|---|---|---|
| 61 | 2200 | $x_3 < R$ | $\overline{y_3}, x_3, y_3, \overline{x_3}$ | $\overline{y_1}, x_3, y_1, \overline{x_3}$ | $S_{segm}(F)$ | $S_{segm}(F´)$ | $\dfrac{(x_3 - \overline{y_3}) * (y_3 - \overline{x_3})}{2}$ | $\dfrac{(x_3 - \overline{x_3}) * (\overline{x_3} - y_1)}{2}$ | $Size^2$ |
| 62 | 2002 | $y_3 < R$ | $x_1, \overline{y_3}, \overline{x_1}, y_3$ | $x_3, \overline{y_3}, \overline{x_3}, y_3$ | $S_{segm}(F)$ | $S_{segm}(F´)$ | $\dfrac{(y_3 - \overline{x_3}) * (x_3 - \overline{y_3})}{2}$ | $\dfrac{(y_3 - \overline{x_3}) * (x_3 - \overline{y_3})}{2}$ | $Size^2$ |
| 63 | 0022 | $|x_1| < R$ | $\overline{y_1}, x_1, \overline{x_1}, y_1$ | $\overline{y_3}, x_1, y_3, \overline{x_1}$ | $S_{segm}(F)$ | $S_{segm}(F´)$ | $\dfrac{(\overline{y_1} - x_1) * (\overline{x_1} - y_1)}{2}$ | $\dfrac{(\overline{y_3} - x_1) * (y_3 - \overline{x_1})}{2}$ | $Size^2$ |
| 64 | 0220 | $|y_1| < R$ | $x_3, \overline{y_1}, \overline{x_3}, y_1$ | $\overline{y_1}, x_1, \overline{x_1}, y_1$ | $S_{segm}(F)$ | $S_{segm}(F´)$ | $\dfrac{(\overline{x_3} - y_1) * (x_3 - \overline{y_1})}{2}$ | $\dfrac{(\overline{x_1} - y_1) * (\overline{y_1} - x_1)}{2}$ | $Size^2$ |

### 6.7. One vertex inside, two vertices on a circle, one vertex outside

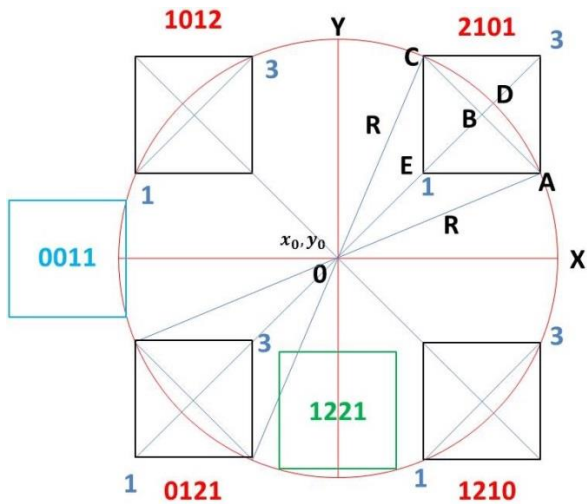

$$S_{intersection} = S_{segmABCD} + S_{triangleAEC}$$

$$S_{segmABCD} = S_{segm}(AC)$$

$$S_{triangleAEC} = \frac{AE * EC}{2} = \frac{Size^2}{2}$$

Fig.12. Location "One vertex inside, two vertices on a circle, one vertex outside"

Obviously, if two adjacent vertices of the square lie on the circumference (blue 0011 and green 1221), then the remaining two of its vertices lie either inside or outside the circle. Thus, not all combinations correspond to some real locations. In below given list of all possible combinations

0112, **0121**, 0211, **1012**, 1021, 1102, 1120, 1201, **1210**, 2011, **2101**, 2110

only 4 of them have not adjacent "1". Therefore, all possible locations are presented in the picture.

As AC is diagonal of square, then its length = Size * $\sqrt{2}$

**Table.10. Conditions and Arguments for the Location "One vertex inside, two vertices on a circle, one vertex outside"**

| | Location | AC | $S_{triangleAEC}$ |
|---|---|---|---|
| 1 | 2101 | Size * $\sqrt{2}$ | $\dfrac{Size^2}{2}$ |
| 2 | 1210 | | |
| 3 | 0121 | | |
| 4 | 1012 | | |

### 6.8. Two vertices inside, one vertex on a circle, one vertex outside

There are many locations in this case with equal description but with different intersections. Therefore, first of all we exclude impossible combinations and then consider each possible one using a separate picture.

All combinations look like:

0122 **0212** 0221 1022 **1202** 1220 2012 **2021** 2102 **2120** 2201 2210

We show that combinations **0212, 1202, 2021 and 2120** correspond to no locations. Recall formulae (12 - 16).

**Case 0212**

$Z_1 > P$ $\qquad\qquad$ $Z_2 < P$
$Z_3 = P$ $\qquad\qquad$ $Z_4 < P$

$Z_1 + Z_3 > 2*P$ $\qquad\qquad$ $Z_2 + Z_4 < 2*P$

As $\quad Z_2 + Z_4 = Z_1 + Z_3 < 2*P$
and $\qquad\quad Z_1 + Z_3 > 2*P$

Latter contradiction means, that combination 0212 corresponds to no location.

**Case 1202**

$Z_1 = P$ $\qquad\qquad$ $Z_2 < P$
$Z_3 > P$ $\qquad\qquad$ $Z_4 < P$

$Z_1 + Z_3 > 2*P$ $\qquad\qquad$ $Z_2 + Z_4 < 2*P$

As $\quad Z_2 + Z_4 = Z_1 + Z_3 < 2*P$
and $\qquad\quad Z_1 + Z_3 > 2*P$

Latter contradiction means, that combination 1202 corresponds to no location.

**Case 2021**

$Z_1 < P$ $\qquad\qquad$ $Z_2 > P$
$Z_3 < P$ $\qquad\qquad$ $Z_4 = P$

$Z_1 + Z_3 < 2*P$ $\qquad\qquad$ $Z_2 + Z_4 > 2*P$

As $\quad Z_2 + Z_4 = Z_1 + Z_3 > 2*P$
and $\qquad\quad Z_1 + Z_3 < 2*P$
Latter contradiction means, that combination 2021 corresponds to no location.

**Case 2120**

$Z_1 < P$ $\qquad\qquad$ $Z_2 = P$
$Z_3 < P$ $\qquad\qquad$ $Z_4 > P$

$Z_1 + Z_3 < 2*P$ $\qquad\qquad$ $Z_2 + Z_4 > 2*P$

As $\quad Z_2 + Z_4 = Z_1 + Z_3 > 2*P$
and $\quad\quad\quad\quad Z_1 + Z_3 < 2*P$
Latter contradiction means, that combination 2120 corresponds to no location.

Now we analyze possible locations. As always we outline various locations of squares with color to simplify the final collection.

### 6.8.1. Location 0122

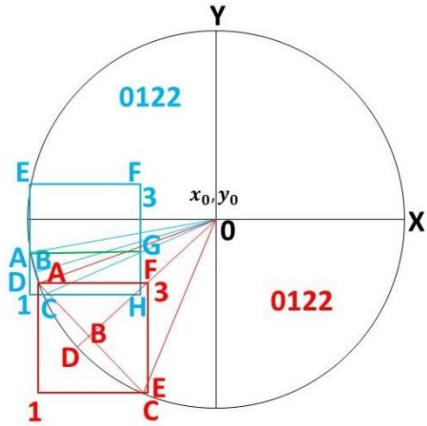

Two possible placing are given in blue and red colors. Corresponding formulae for squares look like

**Blue**

$S_{intersection} = S_{segmABCD} + S_{rectangleAEFG} + S_{trapezoidAGHC}$

**Red**

$S_{intersection} = S_{segmABCD} + S_{trapezoidACEF}$

Fig.13. Location "Two vertices inside, one vertex on a circle, one vertex outside"

As for segment in both cases we use our usual formula $\quad S_{segmABCD} = S_{segm}(AC)$
Now compute parts of formula separately for blue and red cases.

**Blue case**
$S_{rectangleAEFG} = AE * AG = |y_a - y_e| * \text{Size} = 2 * |y_e| * \text{Size} = 2 * |y_3| * \text{Size}$
$S_{trapezoidAGHC} = \frac{AG + HC}{2} * GH = \frac{\text{Size} + HC}{2} * GH = \frac{\text{Size} + |x_h - x_c|}{2} * |y_g - y_h| = \frac{\text{Size} + |x_3 - \overline{y_c}|}{2} * |y_a - y_1| =$
$= \frac{\text{Size} + |x_3 - \overline{y_1}|}{2} * |-y_e - y_1| = \frac{\text{Size} + |x_3 - \overline{y_1}|}{2} * |y_3 + y_1|$
$\mathbf{S_{trapezoidAGHC}} = \frac{\text{Size} + |x_3 - \overline{y_1}|}{2} * |y_3 + y_1|$
$\mathbf{S_{rectangleAEFG}} = 2 * |y_3| * \text{Size}$
If $y_3 = 0 \quad$ then
$S_{trapezoidAGHC} = \frac{\text{Size} + |x_3 - \overline{y_1}|}{2} * |y_3 + y_1| = \frac{\text{Size} + |R - \text{Size} - \overline{\text{Size}}|}{2} * |0 - \text{Size}| = \frac{R - \overline{\text{Size}}}{2} * |\text{Size}| = \frac{\text{Size}*(R - \overline{\text{Size}})}{2}$
$S_{trapezoidAGHC} = \frac{\text{Size}*(R - \overline{\text{Size}})}{2}$
$S_{rectangleAEFG} = 0$

AC = $F(x_1, \overline{y_c}, \overline{x_a}, y_1) = F(x_1, \overline{y_1}, \overline{x_1}, y_1)$

**Red case**
$S_{trapezoidACEF} = \frac{AF + CE}{2} * EF = \frac{\text{Size} + CE}{2} * \text{Size} = \frac{\text{Size} + |x_c - x_e|}{2} * \text{Size} = \frac{\text{Size} + |\overline{y_c} - x_3|}{2} * \text{Size} = \frac{\text{Size} + |\overline{y_1} - x_3|}{2} * \text{Size}$
$\mathbf{S_{trapezoidACEF}} = \frac{\text{Size} + |\overline{y_1} - x_3|}{2} * \text{Size}$
AC = $F(x_1, \overline{y_c}, y_3, y_1) = F(x_1, \overline{y_1}, y_3, y_1)$

We get blue case if $\quad y_3 \geq 0 \quad\quad$ otherwise it is a red case.

### 6.8.2. Location 0221

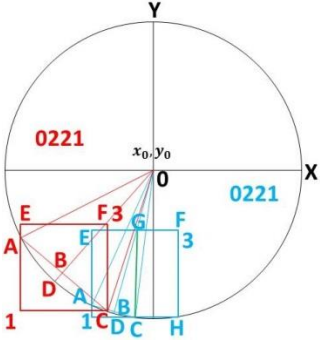

Two possible placing are given in blue and red colors.
Corresponding formulae for squares look like

**Blue**

$$S_{intersection} = S_{segmABCD} + S_{rectangleCGFH} + S_{trapezoidAEGC}$$

**Red**

$$S_{intersection} = S_{segmABCD} + S_{trapezoidAEFC}$$

Fig.14. Location "Two vertices inside, one vertex on a circle, one vertex outside"

As for segment in both cases we use our usual formula $\quad S_{segmABCD} = S_{segm}(AC)$
Now compute parts of formula separately for blue and red cases.

#### Blue case

$S_{rectangleCGFH}$ = CH * CG = $|x_c - x_h|$ * Size = $2 * |x_h|$ * Size = $2 * |x_3|$ * Size

$S_{trapezoidAEGC} = \frac{AE + GC}{2} * GE = \frac{Size + |y_a - y_e|}{2} * |x_g - x_e| = \frac{Size + |\overline{x_a} - y_3|}{2} * |x_c - x_1| = \frac{Size + |\overline{x_1} - y_3|}{2} * |-x_3 - x_1| =$

$= \frac{Size + |\overline{x_1} - y_3|}{2} * |x_3 + x_1|$

$S_{trapezoidAEHC} = \frac{Size + |\overline{x_1} - y_3|}{2} * |x_3 + x_1|$

$S_{rectangleCGFH} = 2 * |x_3|$ * Size

If $x_3 = 0 \quad$ then

$S_{trapezoidAEGC} = \frac{Size + |\overline{x_1} - y_3|}{2} * |x_3 + x_1| = \frac{Size + |R - Size - \overline{Size}|}{2} * |0 - Size| = \frac{R - \overline{Size}}{2} * |Size| = \frac{Size*(R - \overline{Size})}{2}$

$S_{trapezoidAEHC} = \frac{Size*(R - \overline{Size})}{2}$

$S_{rectangleCGFH} = 0$

AC = $F(x_1, \overline{y_c}, \overline{x_a}, y_1)$ = $F(x_1, \overline{y_c}, \overline{x_a}, y_1)$

#### Red case

$S_{trapezoidAEFC} = \frac{AE + FC}{2} * EF = \frac{Size + AE}{2} * Size = \frac{Size + |y_a - y_e|}{2} * Size = \frac{Size + |\overline{x_a} - y_3|}{2} * Size = \frac{Size + |\overline{x_1} - y_3|}{2} * Size$

$S_{trapezoidAEFC} = \frac{Size + |\overline{x_1} - y_3|}{2} * Size$

AC = $F(x_1, x_3, \overline{x_a}, y_1)$ = $F(x_1, x_3, \overline{x_1}, y_1)$

We get blue case if $\quad x_3 \geq 0 \quad\quad$ otherwise it is a red case.

### 6.8.3. Location 1022

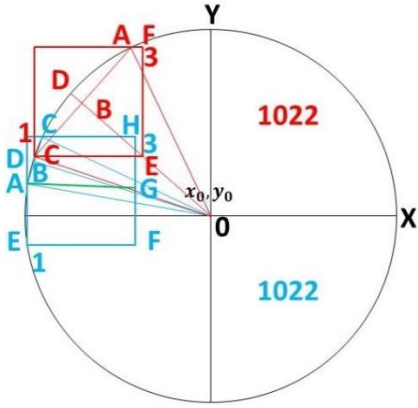

Two possible placing are given in blue and red colors.
Corresponding formulae for squares look like

**Blue**

$$S_{intersection} = S_{segmABCD} + S_{rectangleAGFE} + S_{trapezoidAGHC}$$

**Red**

$$S_{intersection} = S_{segmABCD} + S_{trapezoidAFEC}$$

Fig.15. Location "Two vertices inside, one vertex on a circle, one vertex outside"

As for segment in both cases we use our usual formula $\quad S_{segmABCD} = S_{segm}(AC)$
Now compute parts of formula separately for blue and red cases.

**Blue case**

$$S_{rectangleAGFE} = AE * AG = |y_a - y_e| * \text{Size} = 2 * |y_e| * \text{Size} = 2 * |y_1| * \text{Size}$$

$$S_{trapezoidAGHC} = \frac{AG + HC}{2} * GH = \frac{\text{Size} + HC}{2} * GH = \frac{\text{Size} + |x_h - x_c|}{2} * |y_g - y_h| = \frac{\text{Size} + |x_3 - \overline{y_c}|}{2} * |y_a - y_3| =$$
$$= \frac{\text{Size} + |x_3 - \overline{y_3}|}{2} * |-y_1 - y_3| = \frac{\text{Size} + |x_3 - \overline{y_3}|}{2} * |y_1 + y_3|$$

$$\mathbf{S_{trapezoidAGHC}} = \frac{\text{Size} + |x_3 - \overline{y_3}|}{2} * |y_1 + y_3|$$
$$\mathbf{S_{rectangleAGFE}} = 2 * |y_1| * \text{Size}$$
If $y_1 = 0 \qquad$ then
$$S_{trapezoidAGHC} = \frac{\text{Size} + |x_3 - \overline{y_3}|}{2} * |y_1 + y_3| = \frac{\text{Size} + |R - \text{Size} - \overline{\text{Size}}|}{2} * |0 - \text{Size}| = \frac{R - \overline{\text{Size}}}{2} * |\text{Size}| = \frac{\text{Size} * (R - \overline{\text{Size}})}{2}$$
$$S_{trapezoidAGHC} = \frac{\text{Size} * (R - \overline{\text{Size}})}{2}$$
$$S_{rectangleAGFE} = 0$$

AC = $F(x_1, \overline{y_c}, x_a, y_3) = F(x_1, \overline{y_c}, x_a, y_3)$

**Red case**

$$S_{trapezoidAFEC} = \frac{AF + CE}{2} * EF = \frac{\text{Size} + AF}{2} * \text{Size} = \frac{\text{Size} + |x_a - x_f|}{2} * \text{Size} = \frac{\text{Size} + |\overline{y_a} - x_3|}{2} * \text{Size} = \frac{\text{Size} + |\overline{y_3} - x_3|}{2} * \text{Size}$$
$$\mathbf{S_{trapezoidAFEC}} = \frac{\text{Size} + |\overline{y_3} - x_3|}{2} * \text{Size}$$

AC = $F(\overline{y_a}, x_1, y_3, y_1) = F(\overline{y_a}, x_1, y_3, y_1)$

We get blue case if $\quad y_1 \leq 0 \qquad$ otherwise it is a red case.

### 6.8.4. Location 1220

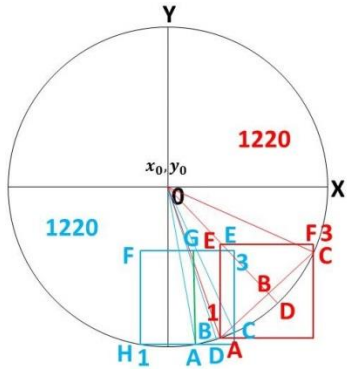

Two possible placing are given in blue and red colors.
Corresponding formulae for squares look like

**Blue**

$$S_{intersection} = S_{segmABCD} + S_{rectangleAGFH} + S_{trapezoidAGEC}$$

**Red**

$$S_{intersection} = S_{segmABCD} + S_{trapezoidAEFC}$$

Fig.16. Location "Two vertices inside, one vertex on a circle, one vertex outside"

As for segment in both cases we use our usual formula $\quad S_{segmABCD} = S_{segm}(AC)$
Now compute parts of formulae separately for blue and red cases.

#### Blue case

$S_{rectangleAGFH} = AH * AG = |x_a - x_h| * Size = 2 * |x_h| * Size = 2 * |x_1| * Size$

$S_{trapezoidAGEC} = \frac{AG + EC}{2} * GE = \frac{Size + |y_e - y_c|}{2} * |x_g - x_e| = \frac{Size + |y_3 - \overline{x_c}|}{2} * |x_a - x_3| = \frac{Size + |y_3 - \overline{x_3}|}{2} * |-x_1 - x_3| = \frac{Size + |y_3 - \overline{x_3}|}{2} * |x_1 + x_3|$

$S_{trapezoidAGEC} = \frac{Size + |y_3 - \overline{x_3}|}{2} * |x_1 + x_3|$

$S_{rectangleAGFH} = 2 * |x_1| * Size$

If $x_1 = 0$ then

$S_{trapezoidAGEC} = \frac{Size + |y_3 - \overline{x_3}|}{2} * |x_1 + x_3| = \frac{Size + |R - Size - \overline{Size}|}{2} * |x_3| = \frac{Size + R - Size - \overline{Size}}{2} * |Size| = \frac{Size * (R - \overline{Size})}{2}$

$S_{trapezoidAGEC} = \frac{Size * (R - \overline{Size})}{2}$

$S_{rectangleAGFH} = 0$

$AC = F(\overline{y_a}, x_3, y_1, \overline{x_c}) = F(\overline{y_1}, x_3, y_1, \overline{x_3})$

#### Red case

$S_{trapezoidAEFC} = \frac{AE + FC}{2} * EF = \frac{Size + FC}{2} * Size = \frac{Size + |y_f - y_c|}{2} * Size = \frac{Size + |y_3 - \overline{x_c}|}{2} * Size = \frac{Size + |y_3 - \overline{x_3}|}{2} * Size$

$S_{trapezoidAEFC} = \frac{Size + |y_3 - \overline{x_3}|}{2} * Size$

$AC = F(\overline{y_a}, x_3, y_1, \overline{x_c}) = F(\overline{y_1}, x_3, y_1, \overline{x_3})$

We get blue case if $\quad x_1 \leq 0 \quad$ otherwise it is a red case.

### 6.8.5. Location 2012

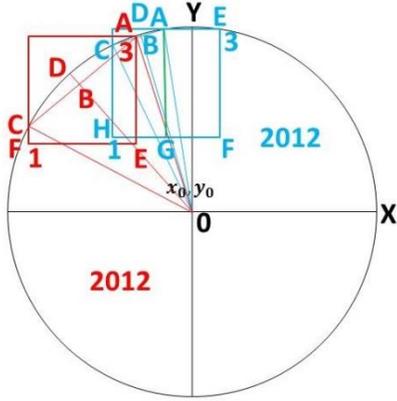

Two possible placing are given in blue and red colors.
Corresponding formulae for squares look like

**Blue**

$$S_{intersection} = S_{segmABCD} + S_{rectangleAGFE} + S_{trapezoidAGHC}$$

**Red**

$$S_{intersection} = S_{segmABCD} + S_{trapezoidAEFC}$$

Fig.17. Location "Two vertices inside, one vertex on a circle, one vertex outside"

As for segment in both cases we use our usual formula $\quad S_{segmABCD} = S_{segm}(AC)$
Now compute parts of formulae separately for blue and red cases.

#### Blue case

$S_{rectangleAGFE}$ = AE * AG = $|x_a - x_e|$ * Size = 2 * $|x_e|$ * Size = 2 * $|x_3|$ * Size

$S_{trapezoidAGHC} = \frac{AG + HC}{2} * GH = \frac{Size + HC}{2} * GH = \frac{Size + |y_h - y_c|}{2} * |x_g - x_h| = \frac{Size + |y_1 - \overline{x_c}|}{2} * |x_a - x_1| =$

$= \frac{Size + |y_1 - \overline{x_1}|}{2} * |-x_3 - x_1| = \frac{Size + |y_1 - \overline{x_1}|}{2} * |x_3 + x_1|$

$S_{trapezoidAGHC} = \frac{Size + |y_1 - \overline{x_1}|}{2} * |x_3 + x_1|$

$S_{rectangleAGFE} = 2 * |x_3|$ * Size

If $x_3 = 0$ then

$S_{trapezoidAGHC} = \frac{Size + |y_1 - \overline{x_1}|}{2} * |x_3 + x_1| = \frac{Size + |R - Size - \overline{Size}|}{2} * |0 - Size| = \frac{R - \overline{Size}}{2} * |Size| = \frac{Size * (R - \overline{Size})}{2}$

$S_{trapezoidAGHC} = \frac{Size * (R - \overline{Size})}{2}$

$S_{rectangleAGFE} = 0$

AC = $F(\overline{y_a}, x_1, y_3, \overline{x_c}) = F(\overline{y_3}, x_1, y_3, \overline{x_1})$

#### Red case

$S_{trapezoidAEFC} = \frac{AE + FC}{2} * EF = \frac{Size + FC}{2} * Size = \frac{Size + |y_f - y_c|}{2} * Size = \frac{Size + |y_1 - \overline{x_c}|}{2} * Size = \frac{Size + |y_1 - \overline{x_1}|}{2} * Size$

$S_{trapezoidAEFC} = \frac{Size + |y_1 - \overline{x_1}|}{2} * Size$

AC = $F(x_3, x_1, y_3, \overline{x_c}) = F(x_3, x_1, y_3, \overline{x_1})$

We get blue case if $\quad x_3 \geq 0 \quad$ otherwise it is a red case.

### 6.8.6. Location 2102

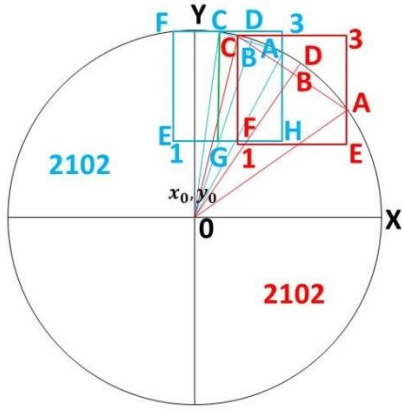

Two possible placing are given in blue and red colors.
Corresponding formulae for squares look like

**Blue**

$S_{intersection} = S_{segmABCD} + S_{rectangleGEFC} + S_{trapezoidAHGC}$

**Red**

$S_{intersection} = S_{segmABCD} + S_{trapezoidAEFC}$

Fig.18. Location "Two vertices inside, one vertex on a circle, one vertex outside"

As for segment in both cases we use our usual formula $\quad S_{segmABCD} = S_{segm}(AC)$
Now compute parts of formulae separately for blue and red cases.

**Blue case**

$S_{rectangleGEFC} = GE * EF = |x_g - x_e| * Size = |x_c - x_1| * Size = |-x_1 - x_1| * Size = 2 * |x_1| * Size$

$S_{trapezoidAHGC} = \frac{AH + GC}{2} * GH = \frac{Size + AH}{2} * GH = \frac{Size + |y_a - y_h|}{2} * |x_g - x_h| = \frac{Size + |\overline{x_a} - y_1|}{2} * |x_c - x_3| =$
$= \frac{Size + |\overline{x_3} - y_1|}{2} * |-x_1 - x_3| = \frac{Size + |\overline{x_3} - y_1|}{2} * |x_1 + x_3|$

$\boldsymbol{S_{trapezoidAHGC}} = \frac{Size + |\overline{x_3} - y_1|}{2} * |x_1 + x_3|$
$\boldsymbol{S_{rectangleGEFC}} = 2 * |x_1| * Size$
If $x_1 = 0$ then
$S_{trapezoidAHGC} = \frac{Size + |\overline{x_3} - y_1|}{2} * |x_1 + x_3| = \frac{Size + |\overline{Size} - (R - Size)|}{2} * |0 - Size| = \frac{Size + |(R - Size) - \overline{Size}|}{2} * |Size| =$
$= \frac{Size + R - Size - \overline{Size}}{2} * |Size| = \frac{R - \overline{Size}}{2} * |Size| = \frac{Size * (R - \overline{Size})}{2}$

$\boldsymbol{S_{trapezoidAHGC}} = \frac{Size * (R - \overline{Size})}{2}$
$S_{rectangleGEFC} = 0$

AC = $F(x_3, \overline{y_c}, \overline{x_a}, y_3) = F(x_3, \overline{y_3}, \overline{x_3}, y_3)$

**Red case**

$S_{trapezoidAEFC} = \frac{AE + FC}{2} * EF = \frac{Size + AE}{2} * Size = \frac{Size + |y_a - y_e|}{2} * Size = \frac{Size + |\overline{x_a} - y_1|}{2} * Size = \frac{Size + |\overline{x_3} - y_1|}{2} * Size$

$\boldsymbol{S_{trapezoidAEFC}} = \frac{Size + |\overline{x_3} - y_1|}{2} * Size$

AC = $F(x_3, \overline{x_3}, \overline{y_c}, y_3) = F(x_3, \overline{x_3}, \overline{y_3}, y_3)$

We get blue case if $\quad x_1 \leq 0 \quad$ otherwise it is a red case.

### 6.8.7. Location 2201

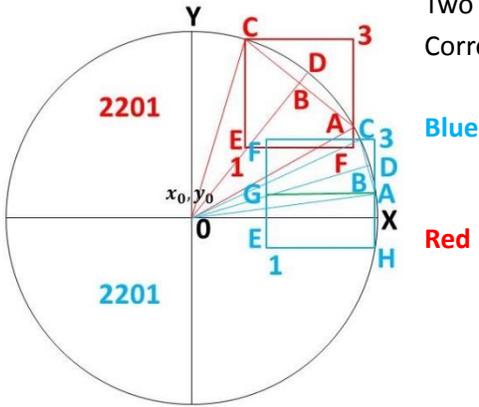

Two possible placing are given in blue and red colors.
Corresponding formulae for squares look like

**Blue**

$S_{intersection} = S_{segmABCD} + S_{rectangleAGEH} + S_{trapezoidACFG}$

**Red**

$S_{intersection} = S_{segmABCD} + S_{trapezoidAFEC}$

Fig.19. Location "Two vertices inside, one vertex on a circle, one vertex outside"

As for segment in both cases we use our usual formula $\quad S_{segmABCD} = S_{segm}(AC)$
Now compute parts of formula separately for blue and red cases.

#### Blue case

$S_{rectangleAGEH} = AH * EH = |y_a - y_h| * Size = |-y_h - y_h| * Size = 2 * |y_1| * Size$

$S_{trapezoidACFG} = \frac{AG + FC}{2} * GF = \frac{Size + |x_f - x_c|}{2} * |y_g - y_f| = \frac{Size + |x_1 - \overline{y_c}|}{2} * |y_a - y_3| = \frac{Size + |x_1 - \overline{y_3}|}{2} * |-y_1 - y_3| =$
$= \frac{Size + |x_1 - \overline{y_3}|}{2} * |y_1 + y_3|$

$S_{trapezoidACFG} = \frac{Size + |\overline{x_3} - y_1|}{2} * |y_1 + y_3|$

$S_{rectangleAGEH} = 2 * |y_1| * Size$

If $y_1 = 0$ then
$S_{trapezoidACFG} = \frac{Size + |x_1 - \overline{y_3}|}{2} * |y_1 + y_3| = \frac{Size + |R - Size - \overline{Size}|}{2} * |0 - Size| = \frac{R - \overline{Size}}{2} * |Size| = \frac{Size * (R - \overline{Size})}{2}$

$S_{trapezoidACFG} = \frac{Size * (R - \overline{Size})}{2}$

$S_{rectangleAGEH} = 0$

AC = $F(x_3, \overline{y_3}, \overline{x_3}, y_3)$

#### Red case

$S_{trapezoidAFEC} = \frac{EC + AF}{2} * EF = \frac{Size + AF}{2} * Size = \frac{Size + |y_a - y_f|}{2} * Size = \frac{Size + |\overline{x_a} - y_1|}{2} * Size = \frac{Size + |\overline{x_3} - y_1|}{2} * Size$

$S_{trapezoidAFEC} = \frac{Size + |\overline{x_3} - y_1|}{2} * Size$

AC = $F(x_3, x_1, \overline{x_3}, y_3)$

We get blue case if following relations hold: $\quad y_1 \leq 0 \quad$ otherwise it is a red case.

### 6.8.8. Location 2210

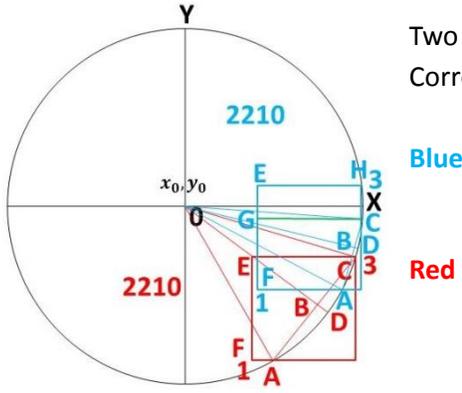

Two possible placing are given in blue and red colors.
Corresponding formulae for squares look like

**Blue**

$$S_{intersection} = S_{segmABCD} + S_{rectangleCHEG} + S_{trapezoidAFGC}$$

**Red**

$$S_{intersection} = S_{segmABCD} + S_{trapezoidAFEC}$$

Fig.20. Location "Two vertices inside, one vertex on a circle, one vertex outside"

As for segment in both cases we use our usual formula $\quad S_{segmABCD} = S_{segm}(AC)$
Now compute parts of formula separately for blue and red cases.

### Blue case
$S_{rectangleCHEG}$ = CH * HE = $|y_c - y_h|$ * Size = 2 * $|y_h|$ * Size = 2 * $|y_3|$ * Size

$S_{trapezoidAFGC} = \frac{CG + AF}{2} * GF = \frac{Size + |x_a - x_f|}{2} * |y_g - y_f| = \frac{Size + |\overline{y_a} - x_1|}{2} * |y_c - y_1| = \frac{Size + |\overline{y_1} - x_1|}{2} * |-y_h - y_1| =$

$= \frac{Size + |\overline{y_1} - x_1|}{2} * |-y_3 - y_1| = \frac{Size + |\overline{y_1} - x_1|}{2} * |y_3 + y_1|$

$\mathbf{S_{trapezoidAFGC}} = \frac{Size + |\overline{y_1} - x_1|}{2} * |y_3 + y_1|$

$\mathbf{S_{rectangleCHEG}} = 2 * |y_3|$ * Size

If $y_3 = 0$ then
$S_{trapezoidAFGC} = \frac{Size + |\overline{Size} - (R - Size)|}{2} * |0 - Size| = \frac{Size + |R - Size - \overline{Size}|}{2} * |Size| = \frac{R - \overline{Size}}{2} * Size = \frac{Size * (R - \overline{Size})}{2}$

$\mathbf{S_{trapezoidAFGC}} = \frac{Size * (R - \overline{Size})}{2}$

$\mathbf{S_{rectangleCHEG} = 0}$

AC = $F(\overline{y_1}, x_3, y_1, \overline{x_3})$

### Red case
AC = $F(x_a, x_c, y_a, y_c) = F(\overline{y_a}, x_3, y_1, y_3) = F(\overline{y_1}, x_3, y_1, y_3)$

$S_{trapezoidAFEC} = \frac{EC + AF}{2} * EF = \frac{Size + AF}{2} * Size = \frac{Size + |x_a - x_f|}{2} * Size = \frac{Size + |\overline{y_a} - x_1|}{2} * Size = \frac{Size + |\overline{y_1} - x_1|}{2} * Size$

$\mathbf{S_{trapezoidAFEC}} = \frac{Size + |\overline{y_1} - x_1|}{2} * Size$

AC = $F(\overline{y_1}, x_3, y_1, y_3)$

We get blue case if $\quad y_3 \geq 0 \quad$ otherwise it is red case.

## Summary of the Locations "Two vertices inside, one vertex on a circle, one vertex outside"

All above considered cases for the location "Two vertices inside, one vertex on a circle, one vertex outside" are collected into below given result table. Each case is presented three times depending on condition to better distinguish various calculus.

Table.11. Conditions and Arguments for the Location "Two vertices inside, one vertex on a circle, one vertex outside"

| | Location | Condition | Arguments of F | $S_{trapezoid}$ | $S_{rectangle}$ |
|---|---|---|---|---|---|
| 1 | 2 | 3 | 4 | 5 | 6 |
| 1 | 0122 | $y_3 > 0$ | $x_1, \overline{y_1}, \overline{x_1}, y_1$ | $\frac{Size + |x_3 - \overline{y_1}|}{2} * |y_3 + y_1|$ | $2 * |y_3| * Size$ |
| 2 | 0122 | $y_3 = 0$ | | $\frac{Size * (R - \overline{Size})}{2}$ | |
| 3 | 0122 | $y_3 < 0$ | $x_1, \overline{y_1}, y_3, y_1$ | $\frac{Size + |\overline{y_1} - x_3|}{2} * Size$ | |
| 4 | 0221 | $x_3 < 0$ | $x_1, \overline{y_c}, \overline{x_a}, y_1$ | $\frac{Size + |\overline{x_1} - y_3|}{2} * |x_3 + x_1|$ | $2 * |x_3| * Size$ |
| 5 | 0221 | $x_3 = 0$ | | $\frac{Size * (R - \overline{Size})}{2}$ | |
| 6 | 0221 | $x_3 > 0$ | $x_1, x_3, \overline{x_1}, y_1$ | $\frac{Size + |\overline{x_1} - y_3|}{2} * Size$ | |
| 7 | 1022 | $y_1 < 0$ | $x_1, \overline{y_c}, x_a, y_3$ | $\frac{Size + |x_3 - \overline{y_3}|}{2} * |y_1 + y_3|$ | $2 * |y_1| * Size$ |
| 8 | 1022 | $y_1 = 0$ | | $\frac{Size * (R - \overline{Size})}{2}$ | |
| 9 | 1022 | $y_1 > 0$ | $\overline{y_a}, x_1, y_3, y_1$ | $\frac{Size + |\overline{y_3} - x_3|}{2} * Size$ | |
| 10 | 1220 | $x_1 < 0$ | $\overline{y_1}, x_3, y_1, \overline{x_3}$ | $\frac{Size + |y_3 - \overline{x_3}|}{2} * |x_1 + x_3|$ | $2 * |x_1| * Size$ |
| 11 | 1220 | $x_1 = 0$ | | $\frac{Size * (R - \overline{Size})}{2}$ | |
| 12 | 1220 | $x_1 > 0$ | $\overline{y_1}, x_3, y_1, \overline{x_3}$ | $\frac{Size + |y_3 - \overline{x_3}|}{2} * Size$ | |
| 13 | 2012 | $x_3 > 0$ | $\overline{y_3}, x_1, y_3, \overline{x_1}$ | $\frac{Size + |y_1 - \overline{x_1}|}{2} * |x_3 + x_1|$ | $2 * |x_3| * Size$ |
| 14 | 2012 | $x_3 = 0$ | | $\frac{Size * (R - \overline{Size})}{2}$ | |
| 15 | 2012 | $x_3 < 0$ | $x_3, x_1, y_3, \overline{x_1}$ | $\frac{Size + |y_1 - \overline{x_1}|}{2} * Size$ | |
| 16 | 2102 | $x_1 < 0$ | $x_3, \overline{y_3}, \overline{x_3}, y_3$ | $\frac{Size + |\overline{x_3} - y_1|}{2} * |x_1 + x_3|$ | $2 * |x_1| * Size$ |
| 17 | 2102 | $x_1 = 0$ | | $\frac{Size * (R - \overline{Size})}{2}$ | |
| 18 | 2102 | $x_1 > 0$ | $x_3, \overline{x_3}, \overline{y_3}, y_3$ | $\frac{Size + |\overline{x_3} - y_1|}{2} * Size$ | |
| 19 | 2201 | $y_1 < 0$ | $x_3, \overline{y_3}, \overline{x_3}, y_3$ | $\frac{Size + |\overline{x_3} - y_1|}{2} * |y_1 + y_3|$ | $2 * |y_1| * Size$ |
| 20 | 2201 | $y_1 = 0$ | | $\frac{Size * (R - \overline{Size})}{2}$ | |
| 21 | 2201 | $y_1 > 0$ | $x_3, x_1, \overline{x_3}, y_3$ | $\frac{Size + |\overline{x_3} - y_1|}{2} * Size$ | |
| 22 | 2210 | $y_3 > 0$ | $\overline{y_1}, x_3, y_1, \overline{x_3}$ | $\frac{Size + |\overline{y_1} - x_1|}{2} * |y_3 + y_1|$ | $2 * |y_3| * Size$ |
| 23 | 2210 | $y_3 = 0$ | | $\frac{Size * (R - \overline{Size})}{2}$ | |
| 24 | 2210 | $y_3 < 0$ | $\overline{y_1}, x_3, y_1, y_3$ | $\frac{Size + |\overline{y_1} - x_1|}{2} * Size$ | |

### 6.9. Three vertices inside, one vertex outside

Four possible locations are outlined in different colors. Corresponding formulae for squares look like

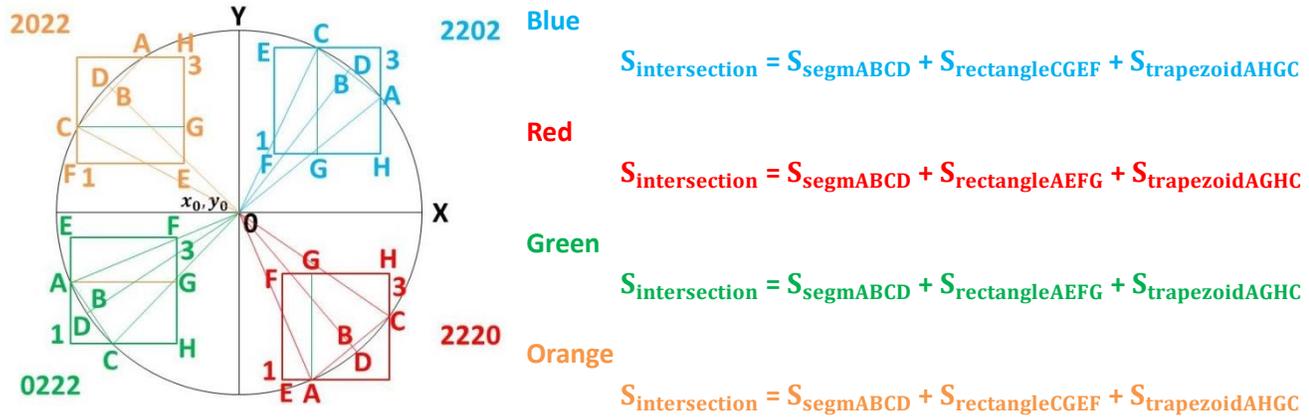

**Blue**
$S_{intersection} = S_{segmABCD} + S_{rectangleCGEF} + S_{trapezoidAHGC}$

**Red**
$S_{intersection} = S_{segmABCD} + S_{rectangleAEFG} + S_{trapezoidAGHC}$

**Green**
$S_{intersection} = S_{segmABCD} + S_{rectangleAEFG} + S_{trapezoidAGHC}$

**Orange**
$S_{intersection} = S_{segmABCD} + S_{rectangleCGEF} + S_{trapezoidAHGC}$

Fig.21. Location "Three vertices inside, one vertex outside"

As for segment in all cases we use our usual formula  $S_{segmABCD} = S_{segm}(AC)$
Now compute parts of formulae separately for each case.

**Blue case**

$S_{rectangleGFEC} = GF * GC = |x_g - x_f| * Size = |x_c - x_1| * Size = |x_c - x_1| * Size = |\overline{y_c} - x_1| * Size = |\overline{y_3} - x_1| * Size$

$S_{trapezoidAHGC} = \frac{GC + AH}{2} * GH = \frac{Size + AH}{2} * GH = \frac{Size + |y_a - y_h|}{2} * |x_g - x_h| = \frac{Size + |y_a - y_h|}{2} * |x_g - x_h| =$

$= \frac{Size + |\overline{x_a} - y_1|}{2} * |x_c - x_3| = \frac{Size + |\overline{x_3} - y_1|}{2} * |\overline{y_c} - x_3| = \frac{Size + |\overline{x_3} - y_1|}{2} * |\overline{y_3} - x_3|$

$S_{trapezoidAHGC} = \frac{Size + |\overline{x_3} - y_1|}{2} * |\overline{y_3} - x_3|$

$S_{rectangleGFEC} = |x_1 - \overline{y_3}| * Size$

$AC = F(x_3, \overline{y_c}, \overline{x_a}, y_3) = F(x_3, \overline{y_3}, \overline{x_3}, y_3)$

**Red case**

$S_{rectangleAEFG} = AE * AG = |x_a - x_e| * Size = |\overline{y_a} - x_1| * Size = |\overline{y_1} - x_1| * Size$

$S_{trapezoidAGHC} = \frac{AG + HC}{2} * GH = \frac{Size + HC}{2} * |x_g - x_h| = \frac{Size + |y_h - y_c|}{2} * |x_a - x_3| = \frac{Size + |y_3 - \overline{x_c}|}{2} * |\overline{y_a} - x_3| =$

$= \frac{Size + |y_3 - \overline{x_3}|}{2} * |\overline{y_1} - x_3|$

$S_{trapezoidAGHC} = \frac{Size + |y_3 - \overline{x_3}|}{2} * |\overline{y_1} - x_3|$

$S_{rectangleAEFG} = |\overline{y_1} - x_1| * Size$

$AC = F(\overline{y_a}, x_3, y_1, \overline{x_c}) = F(\overline{y_a}, x_3, y_1, \overline{x_c})$

**Green case**

$S_{rectangleAEFG} = AE * AG = |y_a - y_e| * Size = |\overline{x_a} - y_f| * Size = |\overline{x_1} - y_3| * Size$

$S_{trapezoidAGHC} = \frac{AG + CH}{2} * GH = \frac{Size + CH}{2} * GH = \frac{Size + |x_c - x_h|}{2} * |y_g - y_h| = \frac{Size + |\overline{y_c} - x_3|}{2} * |y_a - y_1| =$

$= \frac{Size + |\overline{y_1} - x_3|}{2} * |\overline{x_a} - y_1| = \frac{Size + |\overline{y_1} - x_3|}{2} * |\overline{x_1} - y_1|$

$S_{trapezoidAGHC} = \frac{Size + |\overline{y_1} - x_3|}{2} * |\overline{x_1} - y_1|$

$S_{rectangleAEFG} = |\overline{x_1} - y_3| * Size$

$AC = F(x_1, \overline{y_c}, \overline{x_a}, y_1) = F(x_1, \overline{y_1}, \overline{x_1}, y_1)$

**Orange case**

$S_{rectangleCGEF}$ = CF * CG = $|y_c - y_f|$ * Size = $|\overline{x_c} - y_1|$ * Size = $|\overline{x_1} - y_1|$ * Size

$S_{trapezoidAHGC}$ = $\frac{GC + AH}{2}$ * GH = $\frac{Size + AH}{2}$ * $|y_g - y_h|$ = $\frac{Size + |x_a - x_h|}{2}$ * $|y_c - y_3|$ = $\frac{Size + |\overline{y_a} - x_3|}{2}$ * $|\overline{x_c} - y_3|$ =

= $\frac{Size + |\overline{y_3} - x_3|}{2}$ * $|\overline{x_1} - y_3|$

$S_{trapezoidAHGC}$ = $\frac{Size + |\overline{y_3} - x_3|}{2}$ * $|\overline{x_1} - y_3|$

$S_{rectangleCGEF}$ = $|\overline{x_1} - y_1|$ * Size

AC = $F(\overline{y_a}, x_1, y_3, \overline{x_c})$ = $F(\overline{y_3}, x_1, y_3, \overline{x_1})$

Table.12. Conditions and Arguments for the Location "Three vertices inside, one vertex outside"

|   | Location | Arguments of F | $S_{trapezoid}$ | $S_{rectangle}$ |
|---|---|---|---|---|
| 1 | 2202 | $x_3, \overline{y_3}, \overline{x_3}, y_3$ | $\frac{Size + |\overline{x_3} - y_1|}{2}$ * $|\overline{y_3} - x_3|$ | $|x_1 - \overline{y_3}|$ * Size |
| 2 | 2220 | $\overline{y_a}, x_3, y_1, \overline{x_c}$ | $\frac{Size + |y_3 - \overline{x_3}|}{2}$ * $|\overline{y_1} - x_3|$ | $|\overline{y_1} - x_1|$ * Size |
| 3 | 0222 | $x_1, \overline{y_1}, \overline{x_1}, y_1$ | $\frac{Size + |\overline{y_1} - x_3|}{2}$ * $|\overline{x_1} - y_1|$ | $|\overline{x_1} - y_3|$ * Size |
| 4 | 2022 | $\overline{y_3}, x_1, y_3, \overline{x_1}$ | $\frac{Size + |\overline{y_3} - x_3|}{2}$ * $|\overline{x_1} - y_3|$ | $|\overline{x_1} - y_1|$ * Size |

## 7. Summary

In this paper the algorithm was proposed to calculate the area of the intersection of the quadrates of a square mesh with a circle that is arbitrary located on this mesh. For the sake of clarity we have deliberately refused from the evident opportunities of parameterization of presented calculations. Obtained formulae can be directly used for programming.

Below we collect all the formulae from the previous chapters. Each area should be computed as the sum of $S_{segm}$ and values in the columns **6, 7** and **8**. Recall main formulae:

$S_{segmABCD}$ = $S_{segm}(AC)$ = $R^2 * \left[\arcsin\left(\frac{AC}{2R}\right) - \left(\frac{AC}{2R}\right) * \sqrt{1 - \left(\frac{AC}{2R}\right)^2}\right]$

AC = $\sqrt{(x_a - x_c)^2 + (y_a - y_c)^2}$ = $F(x_a, x_c, y_a, y_c)$

$\overline{u}$ = $\sqrt{R^2 - u^2}$

Table.13. Conditions, Arguments and Part Areas for all the Locations. Tally Sheet

| № | Case | Condition | Args of F | AC | $S_{trapezoid}$ | $S_{triangle}$ | $S_{rectangle}$ |
|---|------|-----------|-----------|-----|-----------------|----------------|-----------------|
| 1 | 2 | 3 | 4 | 5 | 6 | 7 | 8 |
| 1 | 0000 | $\|y_1\| < R < \|y_3\|$ | | $2 * \overline{y_1}$ | | | |
| 2 | 0000 | $\|x_1\| < R < \|x_3\|$ | | $2 * \overline{x_1}$ | | | |
| 3 | 0000 | $\|y_1\| < R < \|y_3\|$ | | $2 * \overline{y_3}$ | | | |
| 4 | 0000 | $\|x_3\| < R < \|x_1\|$ | | $2 * \overline{x_3}$ | | | |
| 5 | 1000 | $\|y_1\| < R < \|y_3\|$ | | $2 * \overline{y_1}$ | | | |
| 6 | 0100 | $\|x_1\| < R < \|x_3\|$ | | $2 * \overline{x_1}$ | | | |
| 7 | 0010 | $\|y_3\| < R < \|y_1\|$ | | $2 * \overline{y_3}$ | | | |
| 8 | 0001 | $\|x_3\| < R < \|x_1\|$ | | $2 * \overline{x_3}$ | | | |
| 9 | 2000 | | $\overline{y_1}, x_1, y_1, \overline{x_1}$ | $\sqrt{\|\overline{y_1} - x_1\|^2 + \|\overline{x_1} - y_1\|^2}$ | | $\dfrac{\|\overline{y_1} - x_1\| * \|\overline{x_1} - y_1\|}{2}$ | |
| 10 | 0200 | | $\overline{y_3}, x_1, y_3, \overline{x_1}$ | $\sqrt{\|\overline{x_1} - y_3\|^2 + \|\overline{y_3} - x_1\|^2}$ | | $\dfrac{\|\overline{x_1} - y_3\| * \|\overline{y_3} - x_1\|}{2}$ | |
| 11 | 0020 | | $\overline{y_3}, x_3, y_3, \overline{x_3}$ | $\sqrt{\|\overline{y_3} - x_3\|^2 + \|\overline{x_3} - y_3\|^2}$ | | $\dfrac{\|\overline{y_3} - x_3\| * \|\overline{x_3} - y_3\|}{2}$ | |
| 12 | 0002 | | $\overline{y_1}, x_3, \overline{x_3}, y_1$ | $\sqrt{\|\overline{x_3} - y_1\|^2 + \|\overline{y_1} - x_3\|^2}$ | | $\dfrac{\|\overline{x_3} - y_1\| * \|\overline{y_1} - x_3\|}{2}$ | |
| 13 | 1100 | | | | | | |
| 14 | 0110 | | | Size | | | |
| 15 | 0011 | | | | | | |
| 16 | 1001 | | | | | | |
| 17 | 2100 | | $\overline{y_1}, x_1, y_1, y_3$ | $\sqrt{\|\overline{y_1} - x_1\|^2 + Size^2}$ | | $\dfrac{\|\overline{y_1} - x_1\| * Size}{2}$ | |
| 18 | 1200 | | $x_1, \overline{y_3}, y_1, y_3$ | $\sqrt{\|\overline{y_3} - x_1\|^2 + Size^2}$ | | $\dfrac{\|\overline{y_3} - x_1\| * Size}{2}$ | |
| 19 | 0210 | | $x_1, x_3, \overline{x_1}, y_3$ | $\sqrt{\|\overline{x_3} - y_3\|^2 + Size^2}$ | | $\dfrac{\|\overline{x_3} - y_3\| * Size}{2}$ | |
| 20 | 0120 | | $x_1, x_3, y_3, \overline{x_3}$ | $\sqrt{\|\overline{x_3} - y_1\|^2 + Size^2}$ | | $\dfrac{\|\overline{x_3} - y_1\| * Size}{2}$ | |
| 21 | 0021 | | $\overline{y_3}, x_3, y_3, y_1$ | $\sqrt{\|\overline{y_3} - x_3\|^2 + Size^2}$ | | $\dfrac{\|\overline{y_3} - x_3\| * Size}{2}$ | |
| 22 | 0012 | | $x_3, \overline{y_1}, y_3, y_1$ | $\sqrt{\|\overline{y_1} - x_3\|^2 + Size^2}$ | | $\dfrac{\|\overline{y_1} - x_3\| * Size}{2}$ | |
| 23 | 1002 | | $x_3, x_1, \overline{x_3}, y_1$ | $\sqrt{\|\overline{x_3} - y_1\|^2 + Size^2}$ | | $\dfrac{\|\overline{x_3} - y_1\| * Size}{2}$ | |
| 24 | 2001 | | $x_3, x_1, y_1, \overline{x_1}$ | $\sqrt{\|\overline{x_1} - y_1\|^2 + Size^2}$ | | $\dfrac{\|\overline{x_1} - y_1\| * Size}{2}$ | |
| 25 | 2200 | | $\overline{y_1}, \overline{y_3}, y_1, y_3$ | $\sqrt{(\overline{y_1} - \overline{y_3})^2 + (y_1 - y_3)^2}$ | $(\|\overline{y_1} - x_1\| + \|\overline{y_3} - x_1\|) * \dfrac{Size}{2}$ | | |
| 26 | 0220 | | $x_1, x_3, \overline{x_1}, \overline{x_3}$ | $\sqrt{(x_1 - x_3)^2 + (\overline{x_1} - \overline{x_3})^2}$ | $(\|\overline{x_1} - y_3\| + \|\overline{x_3} - y_3\|) * \dfrac{Size}{2}$ | | |
| 27 | 0022 | | $\overline{y_3}, \overline{y_1}, y_3, y_1$ | $\sqrt{(\overline{y_3} - \overline{y_1})^2 + (y_3 - y_1)^2}$ | $(\|\overline{y_3} - x_3\| + \|\overline{y_1} - x_3\|) * \dfrac{Size}{2}$ | | |
| 28 | 2002 | | $x_3, x_1, \overline{x_3}, \overline{x_1}$ | $\sqrt{(x_3 - x_1)^2 + (\overline{x_3} - \overline{x_1})^2}$ | $(\|\overline{x_3} - y_1\| + \|\overline{x_1} - y_1\|) * \dfrac{Size}{2}$ | | |
| 29 | 2101 | | | | | | |
| 30 | 1210 | | | $Size * \sqrt{2}$ | | $\dfrac{Size^2}{2}$ | |
| 31 | 0121 | | | | | | |
| 32 | 1012 | | | | | | |

**Table.14. Conditions, Arguments and Part Areas for all the Locations. Tally Sheet Continuation**

| № | Case | Condition | Args of F | AC | $S_{trapezoid}$ | $S_{triangle}$ | $S_{rectangle}$ |
|---|---|---|---|---|---|---|---|
| 1 | 2 | 3 | 4 | 5 | 6 | 7 | 8 |
| 33 | 0122 | $y_3 > 0$ | $x_1, \overline{y_1}, \overline{x_1}, y_1$ | | $\frac{Size + |x_3 - \overline{y_1}|}{2} * |y_3 + y_1|$ | | $2 * |y_3| * Size$ |
| 34 | 0122 | $y_3 = 0$ | | | $\frac{Size * (R - \overline{Size})}{2}$ | | |
| 35 | 0122 | $y_3 < 0$ | $x_1, \overline{y_1}, y_3, y_1$ | | $\frac{Size + |\overline{y_1} - x_3|}{2} * Size$ | | |
| 36 | 0221 | $x_3 < 0$ | $x_1, \overline{y_c}, \overline{x_a}, y_1$ | | $\frac{Size + |\overline{x_1} - y_3|}{2} * |x_3 + x_1|$ | | $2 * |x_3| * Size$ |
| 37 | 0221 | $x_3 = 0$ | | | $\frac{Size * (R - \overline{Size})}{2}$ | | |
| 38 | 0221 | $x_3 > 0$ | $x_1, x_3, \overline{x_1}, y_1$ | | $\frac{Size + |\overline{x_1} - y_3|}{2} * Size$ | | |
| 39 | 1022 | $y_1 < 0$ | $x_1, \overline{y_c}, x_a, y_3$ | | $\frac{Size + |x_3 - \overline{y_3}|}{2} * |y_1 + y_3|$ | | $2 * |y_1| * Size$ |
| 40 | 1022 | $y_1 = 0$ | | | $\frac{Size * (R - \overline{Size})}{2}$ | | |
| 41 | 1022 | $y_1 > 0$ | $\overline{y_a}, x_1, y_3, y_1$ | | $\frac{Size + |\overline{y_3} - x_3|}{2} * Size$ | | |
| 42 | 1220 | $x_1 < 0$ | $\overline{y_1}, x_3, y_1, \overline{x_3}$ | | $\frac{Size + |y_3 - \overline{x_3}|}{2} * |x_1 + x_3|$ | | $2 * |x_1| * Size$ |
| 43 | 1220 | $x_1 = 0$ | | | $\frac{Size * (R - \overline{Size})}{2}$ | | |
| 44 | 1220 | $x_1 > 0$ | $\overline{y_1}, x_3, y_1, \overline{x_3}$ | | $\frac{Size + |y_3 - \overline{x_3}|}{2} * Size$ | | |
| 45 | 2012 | $x_3 > 0$ | $\overline{y_3}, x_1, y_3, \overline{x_1}$ | | $\frac{Size + |y_1 - \overline{x_1}|}{2} * |x_3 + x_1|$ | | $2 * |x_3| * Size$ |
| 46 | 2012 | $x_3 = 0$ | | | $\frac{Size * (R - \overline{Size})}{2}$ | | |
| 47 | 2012 | $x_3 < 0$ | $x_3, x_1, y_3, \overline{x_1}$ | | $\frac{Size + |y_1 - \overline{x_1}|}{2} * Size$ | | |
| 48 | 2102 | $x_1 < 0$ | $x_3, \overline{y_3}, \overline{x_3}, y_3$ | | $\frac{Size + |\overline{x_3} - y_1|}{2} * |x_1 + x_3|$ | | $2 * |x_1| * Size$ |
| 49 | 2102 | $x_1 = 0$ | | | $\frac{Size * (R - \overline{Size})}{2}$ | | |
| 50 | 2102 | $x_1 > 0$ | $x_3, \overline{x_3}, \overline{y_3}, y_3$ | | $\frac{Size + |\overline{x_3} - y_1|}{2} * Size$ | | |
| 51 | 2201 | $y_1 < 0$ | $x_3, \overline{y_3}, \overline{x_3}, y_3$ | | $\frac{Size + |\overline{x_3} - y_1|}{2} * |y_1 + y_3|$ | | $2 * |y_1| * Size$ |
| 52 | 2201 | $y_1 = 0$ | | | $\frac{Size * (R - \overline{Size})}{2}$ | | |
| 53 | 2201 | $y_1 > 0$ | $x_3, x_1, \overline{x_3}, y_3$ | | $\frac{Size + |\overline{x_3} - y_1|}{2} * Size$ | | |
| 54 | 2210 | $y_3 > 0$ | $\overline{y_1}, x_3, y_1, \overline{x_3}$ | | $\frac{Size + |\overline{y_1} - x_1|}{2} * |y_3 + y_1|$ | | $2 * |y_3| * Size$ |
| 55 | 2210 | $y_3 = 0$ | | | $\frac{Size * (R - \overline{Size})}{2}$ | | |
| 56 | 2210 | $y_3 < 0$ | $\overline{y_1}, x_3, y_1, y_3$ | | $\frac{Size + |\overline{y_1} - x_1|}{2} * Size$ | | |
| 57 | 2202 | | $x_3, \overline{y_3}, \overline{x_3}, y_3$ | | $\frac{Size + |\overline{x_3} - y_1|}{2} * |\overline{y_3} - x_3|$ | | $|x_1 - \overline{y_3}| * Size$ |
| 58 | 2220 | | $\overline{y_a}, x_3, y_1, \overline{x_c}$ | | $\frac{Size + |y_3 - \overline{x_3}|}{2} * |\overline{y_1} - x_3|$ | | $|\overline{y_1} - x_1| * Size$ |
| 59 | 0222 | | $x_1, \overline{y_1}, \overline{x_1}, y_1$ | | $\frac{Size + |\overline{y_1} - x_3|}{2} * |\overline{x_1} - y_1|$ | | $|\overline{x_1} - y_3| * Size$ |
| 60 | 2022 | | $\overline{y_3}, x_1, y_3, \overline{x_1}$ | | $\frac{Size + |\overline{y_3} - x_3|}{2} * |\overline{x_1} - y_3|$ | | $|\overline{x_1} - y_1| * Size$ |

Table.14. Conditions, Arguments and Part Areas for all the Locations. Tally Sheet Continuation. Special Case.

| № | Special Case | Condition | Args of F - AC | Args of F - A´C´ | $S_{segm}(AC)$ | $S_{segm}(A´C´)$ | $S_{triABC}$ | $S_{triA´B´C´}$ | $S_{square}$ |
|---|---|---|---|---|---|---|---|---|---|
| 61 | 2200 | $x_3 < R$ | $\overline{y_3}, x_3, y_3, \overline{x_3}$ | $\overline{y_1}, x_3, y_1, \overline{x_3}$ | $S_{segm}(F)$ | $S_{segm}(F´)$ | $\dfrac{(x_3 - \overline{y_3}) * (y_3 - \overline{x_3})}{2}$ | $\dfrac{(x_3 - \overline{x_3}) * (\overline{x_3} - y_1)}{2}$ | $Size^2$ |
| 62 | 2002 | $y_3 < R$ | $x_1, \overline{y_3}, \overline{x_1}, y_3$ | $x_3, \overline{y_3}, \overline{x_3}, y_3$ | $S_{segm}(F)$ | $S_{segm}(F´)$ | $\dfrac{(y_3 - \overline{x_3}) * (x_3 - \overline{y_3})}{2}$ | $\dfrac{(y_3 - \overline{x_3}) * (x_3 - \overline{y_3})}{2}$ | $Size^2$ |
| 63 | 0022 | $|x_1| < R$ | $\overline{y_1}, x_1, \overline{x_1}, y_1$ | $\overline{y_3}, x_1, y_3, \overline{x_1}$ | $S_{segm}(F)$ | $S_{segm}(F´)$ | $\dfrac{(\overline{y_1} - x_1) * (\overline{x_1} - y_1)}{2}$ | $\dfrac{(\overline{y_3} - x_1) * (y_3 - \overline{x_1})}{2}$ | $Size^2$ |
| 64 | 0220 | $|y_1| < R$ | $x_3, \overline{y_1}, \overline{x_3}, y_1$ | $\overline{y_1}, x_1, \overline{x_1}, y_1$ | $S_{segm}(F)$ | $S_{segm}(F´)$ | $\dfrac{(\overline{x_3} - y_1) * (x_3 - \overline{y_1})}{2}$ | $\dfrac{(\overline{x_1} - y_1) * (\overline{y_1} - x_1)}{2}$ | $Size^2$ |

## 9. Appendix

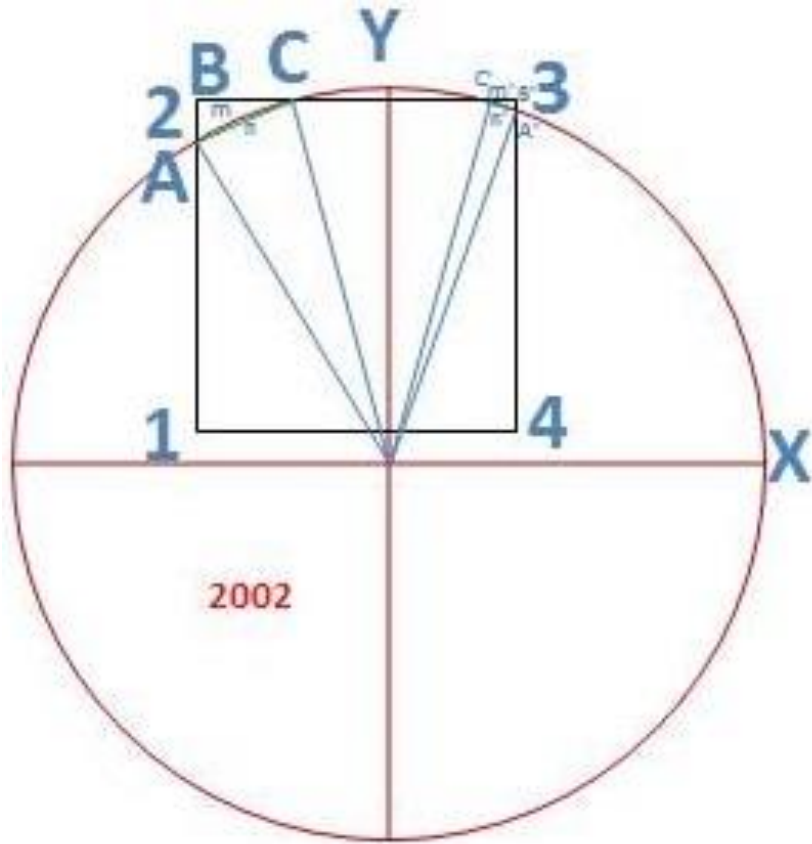

Fig.22. Location "Two vertices inside, two vertices outside. Special case 2002"

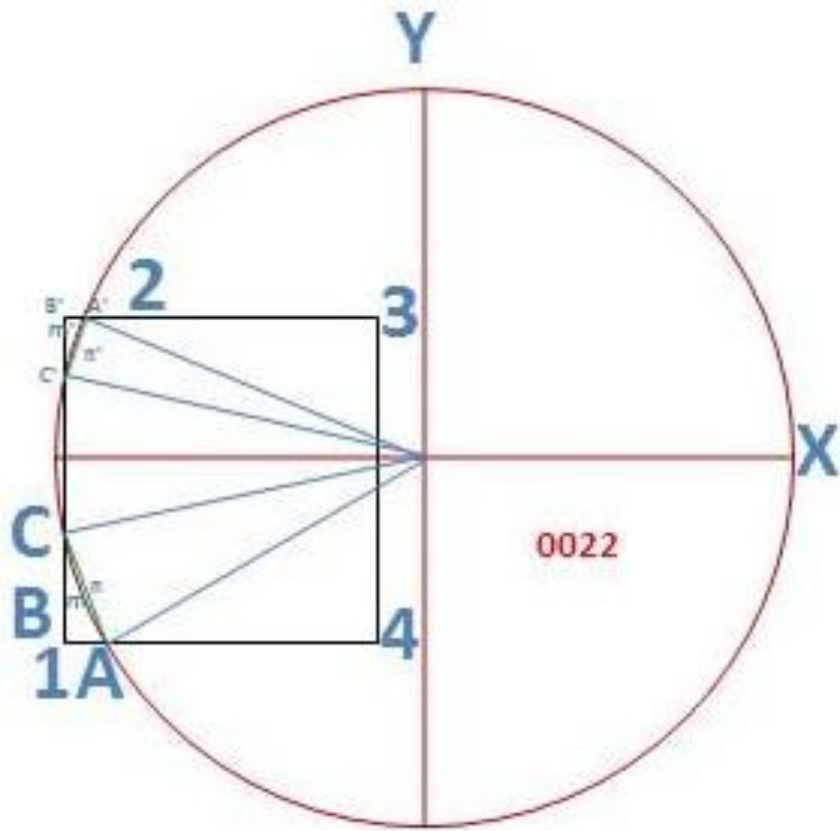

Fig.23. Location "Two vertices inside, two vertices outside. Special case 0022"

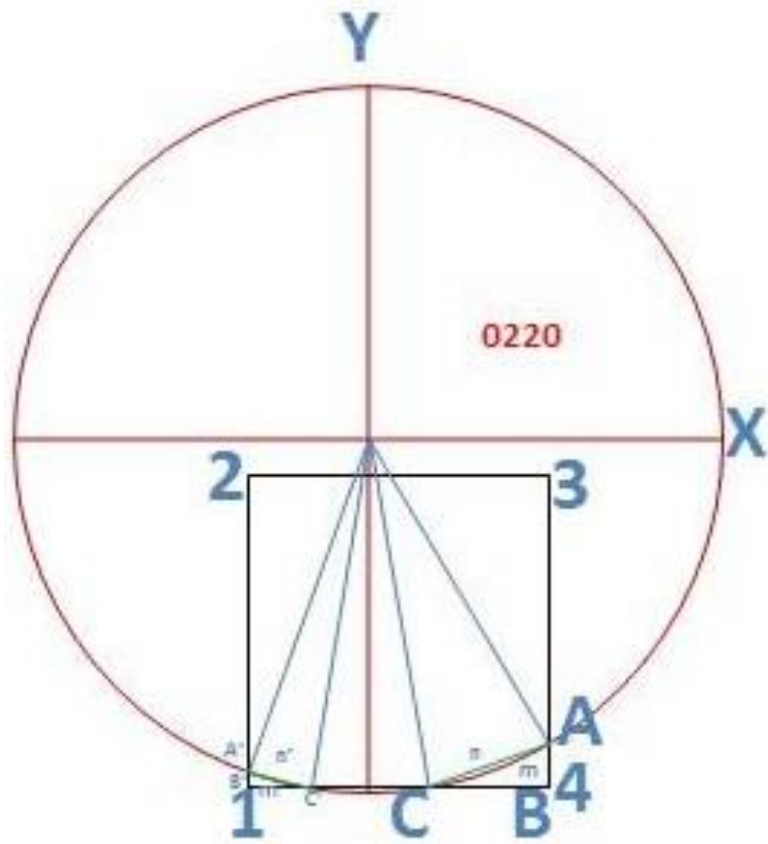

Fig.24. Location "Two vertices inside, two vertices outside. Special case 0220"